\documentclass[manuscript]{aastex}

\def\lax{{$\mathrel{\hbox{\rlap{\hbox{\lower4pt\hbox{$\sim$}}}\hbox{$<$}}}$}}
\def\gax{{$\mathrel{\hbox{\rlap{\hbox{\lower4pt\hbox{$\sim$}}}\hbox{$>$}}}$}}

\usepackage{color}

\usepackage{lscape}
\usepackage{natbib,aas_macros}
\bibliography{apj-jour,reference}
\slugcomment{}

\begin{document}
\title{Neutrino Emissions from Tidal Disruption Remnants}

 \author{
Kimitake \textsc{Hayasaki}\altaffilmark{1}
and
Ryo \textsc{Yamazaki}\altaffilmark{2}
}
\altaffiltext{1}{Department of Astronomy and Space Science, Chungbuk National University, Cheongju 361-763, Korea}
\email{kimi@cbnu.ac.kr}
\altaffiltext{2}{Department of Physics and Mathematics, Aoyama Gakuin University, 5-10-1, Fuchinobe, Sagamihara 252-5258, Japan
}
%%%
%
\begin{abstract}
We study high-energy neutrino emissions from tidal disruption remnants (TDRs) around supermassive black holes. 
The neutrinos are produced by the decay of charged pions originating in ultrarelativistic protons that are accelerated there. In the standard theory of tidal disruption events (TDEs), there are four distinct phases from debris circularization of stellar debris to super- and sub-Eddington to radiatively inefficient accretion flows (RIAFs). In addition, we consider the magnetically arrested disk (MAD) state in both the super-Eddington accretion and RIAF phases. We find that there are three promising cases to produce neutrino emissions: the super-Eddington accretion phase of the MAD state and the RIAF phases of both the non-MAD and MAD states. In the super-Eddington MAD state, the enhanced magnetic field makes it possible to accelerate the protons to $E_{p,{\rm max}}\sim0.35\,{\rm PeV}\,(M_{\rm bh}/10^{7.7}M_\odot)^{41/48}$ with the other given appropriate parameters. The neutrino energy is then $E_{\nu,{\rm pk}}\sim67\,{\rm TeV}\,(M_{\rm bh}/10^{7.7}M_\odot)^{41/48}$ at the peak of the energy spectrum. For $M_{\rm bh}\gtrsim10^{7.7}\,M_\odot$, the neutrino light curve is proportional to $t^{-65/24}$, while it follows the standard $t^{-5/3}$ decay rate for ${M}_{\rm bh}<10^{7.7}\,M_\odot$. In both cases, the large luminosity and characteristic light curves diagnose the super-Eddington MAD state in TDEs. In the RIAF phase of the non-MAD state, we find $E_{p, {\rm max}}\sim0.45\,{\rm PeV}\,(M_{\rm bh}/10^7\,M_\odot)^{5/3}$ and $E_{\nu,{\rm pk}}\sim0.35\,{\rm PeV}\,(M_{\rm bh}/10^7M_\odot)^{5/3}$, and its light curve is proportional to $t^{-10/3}$. This indicates that one can identify whether the existing RIAFs are the TDE origin or not. TDRs are potentially a population of hidden neutrino sources invisible in gamma rays.
\end{abstract}
\keywords{acceleration of particles -- neutrinos -- accretion, accretion disks -- black hole physics -- galaxies: nuclei} 
%
%%%%%%%%%%%
\section{Introduction}
%%%%%%%%%%%
%

% 1st paragraph
%
A recent discovery of very high energy (VHE) neutrinos in the TeV-PeV energy range 
by IceCube \citep{2013PhRvL.111b1103A} has motivated the neutrino astronomy and 
astrophysics. Because the astrophysical neutrinos originate from cosmic-ray hadronic 
interactions, the detection of the neutrinos simultaneously gives us the information about 
the sources of the high-energy cosmic-ray nuclei. There are several astrophysical 
candidates for their origination: active galactic nuclei (AGNs), galaxy clusters/groups, 
starburst galaxies, supernovae and hypernovae, gamma-ray bursts (GRBs), white dwarf (WD) 
mergers, and tidal disruption events (TDEs; for a review see \citealt{2017ARNPS..67...45M}).

%
% 2nd paragraph
%
TDEs are thought to be a key phenomenon in the search for dormant supermassive 
black holes (SMBHs) at the centers of the inactive galaxies and for unidentified 
intermediate-mass black holes (IMBHs) at the centers of star clusters. Most TDEs 
take place when a star at a large separation is perturbed onto a parabolic orbit 
approaching close enough to the SMBH to be ripped apart by its tidal force. The 
subsequent accretion of stellar debris falling back to the SMBH causes a characteristic 
flare with a luminosity large enough to exceed the Eddington luminosity for 
a time scale of weeks to months \citep{rees88,1989ApJ...346L..13E,p89,2009MNRAS.392..332L}. 
Such flares have been discovered at optical \citep{sg+12,ar14,ho14,2016MNRAS.455.2918H,2017ApJ...842...29H}, ultraviolet \citep{2006ApJ...653L..25G,2014ApJ...780...44C,2015ApJ...798...12V}, and 
soft X-ray (\citealt{kb99}; \citealt{2012A&A...541A.106S}; \citealt{2013MNRAS.435.1904M}; 
\citealt{2017ApJ...838..149A}) wavebands with inferred event rates of $10^{-5}-10^{-4}$ 
per year per galaxy \citep{dbeb02,jd04,vf14,sm16}. The other high-energy jetted TDEs 
have been detected through nonthermal emissions in radio \citep{baz+11,2016ApJ...819L..25A,2016Sci...351...62V} 
or hard X-ray \citep{2011Natur.476..421B,2015MNRAS.452.4297B} wavebands with much lower event rates \citep{2014arXiv1411.0704F}. The best observed jetted TDE is Swift J1644+57 \citep{2011Natur.476..421B}; 
others are Swift J2058.4+0516 \citep{2012ApJ...753...77C} and Swift J1112.2-8238 \citep{2015MNRAS.452.4297B}. 
The observed diversity of these optical to X-ray TDEs can be explained in part by the viewing angle of 
the observer relative to the orientation of the disk angular momentum \citep{2018ApJ...859L..20D}.

%
% 3rd paragraph
%
The origin of ultrahigh-energy cosmic rays (UHECRs) is still open to discussion. 
Possible candidates of UHECR accelerators are GRBs \citep{1995PhRvL..75..386W}, 
the short-duration bursts of AGNs \citep{2009ApJ...693..329F}, and so forth. The jetted 
TDEs can also be candidate sources of UHECRs \citep{2014arXiv1411.0704F,2017MNRAS.466.2922P}.
Recent observations with the Pierre Auger Observatory suggest that the compositions 
of UHECRs can be metal rich \citep{2017JCAP...04..038A}. The tidal disruption of a WD 
by an IMBH is proposed as a UHECR source to satisfy 
the heavy nuclei requirement \citep{2017PhRvD..96j3003A,2017PhRvD..96f3007Z}.

%
% 4th paragraph
%
The neutrinos are naturally produced from the UHECRs by pionic decay.
Such high-energy neutrino flux was predicted by \citet{2008AIPC.1065..201M} following 
the scenario of \citet{2009ApJ...693..329F} and for targeting Swift J1644+57 
by \citet{2011PhRvD..84h1301W}. After the detection of the IceCube neutrinos, 
the contribution of the jetted TDEs on the observed neutrino flux was  
examined \citep{2017ApJ...838....3S, 2017PhRvD..95l3001L,2017MNRAS.469.1354D}.
Jetted TDEs can be a population of cosmic-ray accelerators, which are not visible in GeV-TeV gamma-rays, 
that serve as the origin of TeV-PeV neutrinos 
\citep{2016PhRvL.116g1101M,2016PhRvD..93h3005W,2019PhRvD..99f3012M}.
Whether they are a common source of both UHECRs and neutrinos has also been debated
\citep{2018A&A...616A.179G,2018NatSR...810828B}. 
However, little is known about such high-energy emissions from the nonjetted parts including 
the disk components in TDEs, although some sites seem to be good candidates for the production 
of high-energy particles because of the shock formation and the high-energy density around 
the forming disk during the event.

%
% Last paragraph
%
In this work, we examine the stochastic acceleration of the protons by magnetic turbulence 
and subsequent high-energy neutrino emissions from a tidally disrupted star (or a tidal 
disruption remnant (TDR)). In Section~\ref{sec:2}, we discuss the possible sites at which 
the ultrarelativistic protons can be produced during a TDE. In Sectio~\ref{sec:3}, we calculate the energy 
spectral distributions and luminosities of the protons, gamma-rays, and neutrinos 
produced by pionic decay, although the gamma-rays cannot be emitted because 
of the highly opaque remnant. We discuss our results in Section~\ref{sec:dis}. 
Section~\ref{sec:con} is devoted to the conclusion of our scenario.

%
%%%%%%%%%%%%%%%%%%%%%%%%%%%%%%%%%
\section{High-energy emission sites after tidal disruption of a star}
\label{sec:2}
%%%%%%%%%%%%%%%%%%%%%%%%%%%%%%%%%
%

After the tidal disruption of a star, the stellar debris falls back onto an SMBH 
and is circularized by the shock dissipation to convert the orbital energy into 
thermal energy by a collision between the debris head and tail. 
This naturally leads to the formation of an accretion disk around the black hole 
\citep{2013MNRAS.434..909H,2016MNRAS.455.2253B,2016MNRAS.461.3760H}, 
although the detailed dissipation mechanism is still being debated 
\citep{2015ApJ...804...85S,2015ApJ...806..164P}. 
If the accretion rate follows the standard $t^{-5/3}$ decay rate, a TDE can be divided 
by the accretion timescale into the four main evolutionary phases (see equation~\ref{eq:tdeclass}).
In this section, we will discuss the possibility that protons can accelerate 
to the ultrarelativistic energies in each phase.

The tidal disruption radius, $r_{\rm t}$, is given by
\begin{equation}
\frac{r_{\rm t}}{r_{\rm S}}=\left(\frac{M_{\rm bh}}{m_{*}}\right)^{1/3}\frac{r_{*}}{r_{\rm S}}
\approx
5.1\,
\left(\frac{M_{\rm bh}}{10^7\,M_{\odot}}\right)^{-2/3}
\left(\frac{m_{*}}{M_{\odot}}\right)^{-1/3}
\left(\frac{r_{*}}{R_{\odot}}\right),
\end{equation}
where $M_{\rm bh}$ is mass of the central SMBH, $m_*$ and $r_*$ 
are the stellar mass and radius, and $r_{\rm S}=2GM_{\rm bh}/c^2$ is the 
Schwarzschild radius of the SMBH, and $c$ is the speed of light.

This angular momentum conservation allows us to estimate the circularization radius 
of the stellar debris, which is given by 
\begin{eqnarray}
r_{\rm{circ}}
=a_*(1-e_*^2)=(1+e_*)r_{\rm p},
%=\frac{1+e_*}{\beta}r_{\rm{t}},
\label{eq:rc}
\end{eqnarray}
where $a_*$ and $e_*$ are the semi-major axis and the orbital eccentricity of the star, respectively. 
The pericenter distance, $r_{\rm{p}}=a_*(1-e_*)$, is also written by 
\begin{equation}
r_{\rm p}=\frac{r_{\rm t}}{\beta},
\end{equation}
where $\beta$ is the penetration factor, that is, the ratio of tidal disruption to pericenter radii.
The specific binding energy of the stellar debris 
measured at $r_{\rm circ}$ can be then given by
\begin{eqnarray}
\epsilon_{\rm{circ}}=-\frac{1}{2}\frac{1}{1+e_*}\frac{GM_{\rm bh}}{r_{\rm p}}.
\label{eq:ec}
\end{eqnarray}
On the other hand, the specific orbital energy of the star is 
\begin{eqnarray}
\epsilon_*=-\frac{(1-e_*)}{2}\frac{GM_{\rm bh}}{r_{\rm p}}.
\label{eq:eini}
\end{eqnarray} 
The difference between $m_*\epsilon_{*}$ and $m_*\epsilon_{\rm circ}$ gives the maximum amount 
of binding energy potentially dissipated during debris circularization:
\begin{eqnarray}
\Delta\epsilon_{\rm{circ}}
&=&
m_*|\epsilon_{*}-\epsilon_{\rm circ}|
=
\frac{m_*}{2}\frac{{e_{*}^2}}{(1+e_{*})}\frac{GM_{\rm bh}}{r_{\rm p}}
\sim
4.4\times10^{52}\,{\rm{erg}}\,
\nonumber \\
&\times&
\left(\frac{\beta}{1.0}\right)
\left(\frac{m_*}{M_\odot}\right)^{4/3}\left(\frac{r_*}{R_\odot}\right)^{-1}
\left(\frac{M_{\rm{bh}}}{10^7\,M_\odot}\right)^{2/3}
\label{eq:edis}
\end{eqnarray}
for $e_*\approx1.0$ stellar orbits.

Let us assume that the dissipated energy during the debris circularization 
is proportional to the mass fallback rate:
\begin{equation}
L_{\rm circ}=\eta_{\rm circ}\dot{M}_{\rm fb}c^2,
\label{eq:lcirc}
\end{equation}
where $\eta_{\rm circ}$ is the mass-to-energy conversion efficiency of the debris circularization and 
\begin{eqnarray}
\dot{M}_{\rm fb}=\frac{1}{3}\frac{m_{*}}{t_{\rm mtb}}\left(\frac{t}{t_{\rm mtb}}\right)^{-5/3}
\sim5.9\times10^{25}\,{\rm g\,s^{-1}}
\left(\frac{M_{\rm bh}}{10^7\,M_{\odot}}\right)^{-1/2}
\left(\frac{m_{*}}{M_{\odot}}\right)^{2}\left(\frac{r_{*}}{R_{\odot}}\right)^{-3/2}
\label{eq:fbr}
\end{eqnarray}
is the mass fallback rate \citep{1989ApJ...346L..13E}. 
Here $t_{\rm mtb}$ is the orbital period of the stellar debris on the most tightly bound orbit:
\begin{equation}
t_{\rm mtb}=\frac{\pi}{\sqrt{2}}\frac{1}{\Omega_{*}}\left(\frac{M_{\rm bh}}{m_*}\right)^{1/2}
\approx1.1\times10^{7}\,{\rm s}\,
\left(\frac{M_{\rm bh}}{10^7\,M_{\odot}}\right)^{1/2}
\left(\frac{m_{*}}{M_{\odot}}\right)^{-1}\left(\frac{r_{*}}{R_{\odot}}\right)^{3/2},
\label{eq:mtb}
\end{equation}
where $\Omega_{*}=\sqrt{Gm_{*}/r_{*}^{3}}$ is the dynamical angular frequency of the star. 
By using equations (\ref{eq:edis})-(\ref{eq:fbr}), we define the circularization timescale as 
\begin{eqnarray}
{t}_{\rm circ}
\equiv
\frac{\Delta\epsilon_{\rm circ}}{L_{\rm circ}(t_{\rm circ})}
&=&
\left(
\frac{4}{3}\frac{\eta_{\rm circ}}{\beta}\frac{(1+e_{*})}{e_*^2}\frac{r_{\rm t}}{r_{\rm S}}
\right)^{3/2}
t_{\rm mtb}
\nonumber \\
&\sim&
1.8\times10^7\,{\rm s}\,
\left(\frac{\eta_{\rm circ}}{0.1}\right)^{3/2}
\left(\frac{\beta}{1.0}\right)^{-3/2}
\left(\frac{M_{\rm bh}}{10^7\,M_{\odot}}\right)^{-1/2}
\left(\frac{r_{*}}{R_{\odot}}\right)^{3}
\left(\frac{m_{*}}{M_{\odot}}\right)^{-3/2}
\label{eq:deltcirc}
\end{eqnarray}
for $e_*\approx1$.
Because $t_{\rm circ}$ should be longer than $t_{\rm mtb}$ so that 
the debris circularization starts after the most tightly bound debris firstly 
falls back to the black hole, $\eta_{\rm circ}$ should be larger than a 
certain critical value $\eta_0$, which is given by
\begin{eqnarray}
\eta_0
=\frac{3\beta}{4}\frac{e_*^2}{1+e_*}\frac{r_{\rm S}}{r_{\rm t}}
\sim7.4\times10^{-2}
\left(\frac{\beta}{1.0}\right)
\left(\frac{M_{\rm bh}}{10^7\,M_{\odot}}\right)^{2/3}
\left(\frac{r_{*}}{R_{\odot}}\right)^{-1}
\left(\frac{m_{*}}{M_{\odot}}\right)^{1/3}
\end{eqnarray} 
for $e_*\approx1$.
If the debris circularization is done only through the shock dissipation by the debris self-crossings, 
the circularization timescale for nonmagnetized stellar debris can be estimated as 
${t}_{\rm circ}\approx8.3\,(M_{\rm bh}/10^6\,M_\odot)^{-3/5}\beta^{-3}t_{\rm mtb}$ based on the ballistic approximation \citep{2017MNRAS.464.2816B}. Equating this equation with equation (\ref{eq:deltcirc}), we can evaluate the circularization efficiency as
\begin{eqnarray}
\eta_{\rm circ}
\sim1.6
%\left(\frac{8.3}{10^{3/5}}\right)^{2/3}
\left(\frac{1.0}{\beta}\right)^{-2}
\left(
\frac{M_{\rm bh}}{10^7\,M_\odot}
\right)^{-2/5}
\eta_0.
\end{eqnarray}
This is applicable if the black hole mass is less than $\sim3.4\times10^7(\beta/1.0)^{-5}\,M_\odot$ because of $\eta_{\rm circ}>\eta_0$.

Now we define the normalized accretion rate by 
\begin{equation}
\dot{m}\equiv\frac{\dot{M}}{\dot{M}_{\rm Edd}},
\label{eq:mmdot}
\end{equation}
where $\dot{M}_{\rm Edd}=L_{\rm Edd}/c^2\sim1.4\times10^{24}\,{\rm g\,s^{-1}}(M_{\rm bh}/10^7\,M_\odot)$ and 
$L_{\rm Edd}=4\pi{GM_{\rm bh}}m_{p}c/\sigma_{\rm T}$ is the Eddington luminosity, $m_{p}$ 
is the proton mass, and $\sigma_{\rm T}$ is the Thomson scattering cross section. 
We can estimate the time when it takes from super-Eddington to sub-Eddington accretion as
\begin{eqnarray}
t_{\rm Edd}=
\left(\frac{1}{3}
%\frac{1}{\dot{m}}
\frac{m_*}{t_{\rm mtb}}\frac{1}{\dot{M}_{\rm Edd}}\right)^{3/5}t_{\rm mtb}
\sim
1.1\times10^{8}\,{\rm s}\,
\left(\frac{M_{\rm bh}}{10^7\,M_\odot}\right)^{-2/5}
\left(\frac{m_*}{M_\odot}\right)^{1/5}
\left(\frac{r_*}{R_\odot}\right)^{3/5}
\label{eq:tEdd}
\end{eqnarray}
by substituting equation (\ref{eq:fbr}) into equation (\ref{eq:mmdot}) with $\dot{m}=1$. Because $t_{\rm circ}/t_{\rm Edd}\sim
0.17\,(\eta_{\rm circ}/0.1)^{3/2}(\beta/1.0)^{-3/2}\left(M_{\rm bh}/10^7\,M_\odot\right)^{-1/10}
\left(m_*/M_\odot\right)^{-17/10}\left(r_*/{R_\odot}\right)^{12/5}$ is smaller than 1, the 
circularization phase is shorter than the super-Eddington accretion phase. 
As time goes by, $\dot{m}$ decreases to $0.01$, and at this point 
the accretion disk enters the radiatively inefficient accretion flow (RIAF) phase. 
We set a time of onset of the RIAF phase at $\dot{m}=0.01$ as 
\begin{eqnarray}
t_{\rm RIAF}=\dot{m}^{-3/5}t_{\rm Edd}
%\left(\frac{1}{3}\frac{1}{\dot{m}}\frac{m_*}{t_{\rm mtb}}\frac{1}{\dot{M}_{\rm Edd}}\right)^{3/5}t_{\rm mtb}
\sim1.7\times10^{9}\,{\rm s}
\left(\frac{\dot{m}}{0.01}\right)^{-3/5}
\left(\frac{M_{\rm bh}}{10^7\,M_\odot}\right)^{-2/5}
\left(\frac{m_*}{M_\odot}\right)^{1/5}
\left(\frac{r_*}{R_\odot}\right)^{3/5}.
\label{eq:triaf}
\end{eqnarray}

Here we divide the TDR into the four evolutionary phases:
%
%%%%%%%%%
\begin{eqnarray}
\left\{ \begin{array}{ll}
t_{\rm mtb}<t \lesssim t_{\rm circ} & {\verb|Circularization phase|} \\
t_{\rm circ} \lesssim t \lesssim t_{\rm Edd} & {\verb|Super-Eddington accretion phase|} \\
t_{\rm Edd} \lesssim t \lesssim t_{\rm RIAF} & {\verb|Sub-Eddington accretion phase|} \\
t_{\rm RIAF} \lesssim t & {\verb|RIAF phase|}. \\
\end{array} \right.
\label{eq:tdeclass}
\end{eqnarray}
%%%%%%%%%
%
If a star on a marginally hyperbolic orbit is tidally disrupted by an SMBH, 
the RIAF phase would start right after the circularization phase 
without going through both the super-Eddington and sub-Eddington 
accretion phases \citep{2018ApJ...855..129H}.

For the standard disk model, the number density is given by
\begin{equation}
n_p=\frac{\dot{M}}{2\pi{m}_{p}v_{r}r^2},
%\sim3.1\times10^{14}\,({\rm{cm^{-3}}})\,\left(\frac{\dot{m}}{\dot{m}_{\rm fb}}\right)\left(\frac{M_{\rm bh}}{10^6\,M_\odot}\right)^{-1/2}\left(\frac{r_{\rm t}}{r_{\rm S}}\right)^{-5/2},
\label{eq:npst}
\end{equation}
where the radial drift velocity is given by
\begin{equation}
v_r\approx\alpha{v_{\rm K}}\left(\frac{H}{r}\right)^2
\end{equation}
with the Shakura-Sunyaev viscosity parameter $\alpha$, 
Keplerian velocity $v_{\rm K}(r)=\sqrt{GM/r}$, and the disk thickness $H$.
Note that the geometrically thin disk approximation that $H/r\sim0.01$ is adopted for the standard disk.
The accretion time is then given by
\begin{eqnarray}
t_{\rm inf}=\frac{r}{v_r}=\frac{r}{\alpha\,v_{\rm K}}\left(\frac{H}{r}\right)^{-2}\sim1.6\times10^8\,{\rm s}\,
\left(\frac{\alpha}{0.1}\right)
\left(\frac{H/r}{0.01}\right)^{-2}
\left(\frac{M_{\rm bh}}{10^7\,M_\odot}\right)^{-1/2}
\left(\frac{r}{r_{\rm p}}\right)^{3/2}.
\end{eqnarray}
%We use $r$ as the size of the region where the protons can accelerate in what follows. 
The proton-proton relaxation timescale is estimated as
\begin{eqnarray}
t_{\rm rel}&=&\frac{4\sqrt
{\pi}}{\tau_{p}\ln\Lambda}
%\frac{1}{\tau_{p}}
%\frac{1}{n_{p}\sigma_{\rm T}c}
\left(\frac{m_p}{m_e}\right)^2
\left(\frac{k_{\rm B}T}{m_{p}c^2}\right)^{3/2}
\frac{r}{c} \nonumber \\
&\sim&5.7\times10^6\,{\rm s}\,
\left(\frac{\alpha}{0.1}\right)
\left(\frac{20}{\ln\Lambda}\right)
\left(\frac{\dot{m}}{0.1}\right)^{-1}
\left(\frac{M_{\rm bh}}{10^7\,M_\odot}\right)
\left(\frac{r}{r_{\rm p}}\right)^{-3/2},
%\alpha_{-1}M_{6}\dot{m}_{0.01}^{-1},
\label{eq:trelax}
\end{eqnarray}
where $\tau_{p}={n}_p\sigma_{\rm T}r$, $\ln\Lambda$, and $m_{e}$ are the optical depth for Thomson scattering, 
the Coulomb logarithm, and the electron mass, respectively. Here we assume the kinetic energy is completely converted to the thermal energy. The resultant proton's temperature is given by $k_{\rm B}T=(1/3)GM_{\rm bh}/r$.

The Coulomb loss time is given by equation (29) of \cite{1996ApJ...456..106D} as
\begin{eqnarray}
t_{\rm Coul}=\frac{1225}{\tau_p\ln\Lambda}
%\frac{\gamma^2}{\gamma+1}
\frac{(\gamma-1)}{(v_p/c)^2}
\left[3.8\theta_{\rm e}^{3/2}+\left(\frac{v_p}{c}\right)^3
\right]\frac{r}{c},
\label{eq:tcoul}
\end{eqnarray}
where $\gamma=1/\sqrt{1-(v_p/c)^2}$ is the Lorentz factor of the proton with velocity $v_{\rm p}$, 
and $\theta_e=k_{\rm B}T_{e}/(m_{e}c^2)$ is the normalized electron temperature.
Adopting $\gamma\sim10$, $v_{p}/c\approx1$, and $\theta_e\ll{v}_{p}/c$ in equation (\ref{eq:tcoul}), 
we obtain 
\begin{eqnarray}
t_{\rm Coul}&\approx&\frac{1225}{\tau_{p}\ln\Lambda}
\frac{(\gamma-1)\sqrt{\gamma^2-1}}{\gamma}\frac{r}{c} \nonumber \\
&\sim&4.4\times10^5\,{\rm s}\,
\left(\frac{\alpha}{0.1}\right)
\left(\frac{20}{\ln\Lambda}\right)
\left(\frac{\dot{m}}{0.1}\right)^{-1}
\left(\frac{M_{\rm bh}}{10^7\,M_\odot}\right)^{-1/2}
\left(\frac{r}{r_{\rm p}}\right)^{3/2}.
\label{eq:tcoul2}
\end{eqnarray}
Since both $t_{\rm rel}$ and $t_{\rm Coul}$ are clearly shorter than the accretion timescale 
for the typical parameters during the sub-Eddington accretion phase, the plasma is collisional, so that nonthermal, 
high-energy protons are unlikely to be accelerated\footnote{See also panel (c) of Figure~1, where the characteristic 
timescales including these three are compared with the acceleration's timescale.}. 
Therefore, we will consider the other possible sites of the TDR that produce high-energy emissions in the following sections.

%
%%%%%%%%%%%%%%%%%%%%%%%%%%%%%%%%%%%%
\subsection{Collision of stellar debris after the tidal disruption}
%%%%%%%%%%%%%%%%%%%%%%%%%%%%%%%%%%%%
%
% As the pericenter distance is closer to the Schwarzschild radius of the SMBH, 

When the stellar debris passes through the pericenter distance, 
it significantly changes the trajectory of the debris 
by general relativistic apsidal precession \citep{rees88,2013MNRAS.434..909H}. 
According to \cite{2016ApJ...830..125J}, the most tightly bound debris
experiences the relativistic perihelion shift by the angle per orbit, 
\begin{equation}
\omega_{\rm S}=\frac{3\pi}{(1+e_{\rm mtb})}\frac{{r}_{\rm S}}{r_{\rm p}}
\approx\frac{3\pi}{2}\frac{{r}_{\rm S}}{r_{\rm p}}
\label{eq:omegas}
\end{equation}
to the lowest post-Newtonian order \citep{2010PhRvD..81f2002M}, 
where $e_{\rm mtb}\approx1$ is the debris eccentricity of the most tightly bound orbit. 
The radial distance of the stream-stream collision from the SMBH, 
where the debris collides with each other for the first time after the tidal disruption, 
can be written with equation~(\ref{eq:omegas}) by (see \citealt{2015ApJ...812L..39D})
\begin{equation}
r_{\rm c}=\frac{a(1-e_{\rm mtb}^{2})}{(1-e_{\rm mtb}\cos(\omega_{\rm S}/2))}
\approx\frac{16}{\omega_{\rm S}^{2}}r_{\rm p}
%\frac{r_{\rm t}}{\beta}
=\frac{64}{9\pi^{2}}\left(\frac{r_{\rm p}}{r_{\rm S}}\right)^{2}r_{\rm p},
\end{equation}
where we assume that $e_{\rm mtb}\approx1$ and $\omega_{\rm S}\ll\pi$.

Since the stellar debris moves on a highly eccentric orbit, the debris velocity 
at $r_{\rm c}$ is estimated to be
\begin{equation}
v_{\rm c}=v_{\rm K}(r_{\rm c})\approx\frac{3\pi}{8\sqrt{2}}
%\beta^{3/2}
\left(\frac{r_{\rm p}}{r_{\rm S}}\right)^{-3/2}c. 
\label{eq:vc}
\end{equation}
Assuming that the debris stream expands homologously,
the radius of the stream cross section is given by
\begin{equation}
R(r_{\rm c})=\left(\frac{r_{\rm c}}{r_{\rm p}}\right)r_{*}
\approx\frac{64}{9\pi^{2}}
\left(\frac{r_{\rm p}}{r_{\rm S}}\right)^{2}r_{*}
\sim1.3\times10^{12}\,{\rm cm}\,\left(\frac{r_*}{R_\odot}\right)
\left(\frac{r_{\rm p}}{r_{\rm S}}\right)^{2}.
\label{eq:zc}
\end{equation}
The proton's number density at $r_{\rm c}$ is given by
\begin{equation}
n_{p}=\frac{\dot{M}}{m_{p}v_{\rm c}\sigma_{\rm c}}\sim1.6\times10^{15}\,{\rm{cm^{-3}}}\,
\left(\frac{\dot{m}}{\dot{m}_{\rm fb}}\right)
\left(\frac{M_{\rm bh}}{10^6\,M_\odot}\right)^{-1/2}
\left(\frac{r_{\rm p}}{r_{\rm S}}\right)^{-5/2},
\label{eq:np}
\end{equation}
where $\sigma_{c}=\pi{R(r_{\rm c})}^{2}$ is the cross section 
of the return debris and we obtain $\dot{m}_{\rm fb}$ 
\begin{equation}
\dot{m}_{\rm fb}=\frac{\dot{M}_{\rm fb}(t_{\rm mtb})}{\dot{M}_{\rm Edd}}
\sim42\,\left(\frac{M_{\rm bh}}{10^7\,M_\odot}\right)^{-3/2}
\left(\frac{m_*}{M_\odot}\right)^{2}\left(\frac{r_*}{R_\odot}\right)^{-3/2}
\label{eq:mdotfb}
\end{equation}
by using equations (\ref{eq:fbr}), (\ref{eq:mtb}), and (\ref{eq:mmdot}).
%The strength of the magnetic field enhanced by the local dynamo mechanism 

Now we examine whether the first-order particle (Fermi) acceleration is efficient at the shock.    
The mean free path of the photons, during which they can travel until colliding with the protons, is estimated to be 
\begin{equation}
l_{\nu}=\frac{1}{n_{p}\sigma_{\rm T}}\sim9.6\times10^{8}\,{\rm{cm}}\,\left(\frac{\dot{m}}{\dot{m}_{\rm fb}}\right)^{-1}
\left(\frac{M_{\rm bh}}{10^7\,M_\odot}\right)^{1/2}
\left(\frac{r_{\rm p}}{r_{\rm S}}\right)^{5/2}.
\label{eq:mfp}
\end{equation}
The Larmor radius of the proton of $\gamma\,(v_{p}/c)\sim1$ is given by
\begin{equation}
r_{\rm L}\sim\frac{m_{p}c^{2}}{q_{e}B},
%=\frac{m_{p}c^2}{e}\sqrt{\frac{3\mathcal{B}}{8\pi}}\sqrt{\frac{r}{GMm_{p}n_{p}}},
\label{eq:rl}
\end{equation} 
where $q_{e}$ is the electric charge and $B$ is the magnetic field strength. 
It is obtained with the plasma beta $\mathcal{B}$ based on the energy equipartition 
assumption by
\begin{equation}
B=\sqrt{
\frac{8\pi{m_p}n_{p}k_{\rm B}T}{\mathcal{B}}
}
=
\left(\frac{8\pi}{3\mathcal{B}}\right)^{1/2}\sqrt{\rho_{p}v_{\rm K}(r)^2},
\label{eq:bfield}
\end{equation}
where $\rho_{p}=m_{p}n_{p}$ is the proton's mass density. 
The ratio of the mean free path to the Larmor radius at $r_{\rm c}$ is estimated to be
\begin{eqnarray}
\frac{l_{\nu}}{r_{\rm L}}
&=&
\frac{q_{\rm e}B}{n_{p}m_{p}c^{2}\sigma_{\rm T}}
=\frac{q_{\rm e}}{n_{p}m_{p}c^{2}\sigma_{\rm T}}
\left(\frac{8\pi}{3\mathcal{B}}\right)^{1/2}
\sqrt{\frac{GMm_{p}n_{p}}{r_{\rm c}}} 
\sim
5.7\times10^7\,
\left(\frac{\mathcal{B}}{3}\right)^{-1/2}
\nonumber \\
&\times&
\left(\frac{\dot{m}}{\dot{m}_{\rm fb}}\right)^{-1/2}
\left(\frac{M_{\rm bh}}{10^7\,M_\odot}\right)^{1/4}
\left(\frac{m_*}{M_\odot}\right)^{-1}
\left(\frac{r_*}{R_\odot}\right)^{7/4}
\left(\frac{r_{\rm p}}{r_{\rm S}}\right)^{-3/4}
\left(\frac{r_{\rm c}}{r_{\rm p}}\right)^{1/4}.
\label{eq:ltorl}
\end{eqnarray}
It is noted from equations (\ref{eq:zc}), (\ref{eq:mfp}), and (\ref{eq:ltorl}) that $r_{\rm L}\ll{l_\nu}\ll{R(r_{\rm c})}$. 
The radiation-mediated shock should be formed at the first shock of debris circularization, 
leading to the inefficient first-order Fermi acceleration there \citep{2001PhRvL..87g1101W,2013PhRvL.111l1102M}. 

Next we discuss the possibility of causing the second-order Fermi acceleration in the magnetic turbulence 
that is excited during the debris circularization phase. The characteristic timescales for the second-order Fermi 
acceleration of the protons are evaluated by \citep{2015ApJ...806..159K}
\begin{equation}
t_{\rm accl}=\frac{1}{\zeta}\frac{r}{c}\left(\frac{v_{\rm A}}{c}\right)^{-2}\left(\frac{r_{\rm L}}{r}\right)^{2-s}\gamma^{2-s},
\label{eq:taccl}
\end{equation}
where $\zeta$ shows the ratio of the strength of turbulent fields to that of the nonturbulent fields, 
$s$ is a spectral index of the turbulence, and the Alfv\'en speed is calculated as 
\begin{equation}
v_{\rm A}=\frac{B}{\sqrt{4\pi{m_{p}}n_{p}}}
=\left(\frac{2}{3\mathcal{B}}\right)^{1/2}v_{\rm K}(r)=\frac{c}{\sqrt{3\mathcal{B}}}
\left(\frac{r}{r_{\rm S}}\right)^{-1/2}.
\label{eq:alfven}
\end{equation} 
Throughout this paper, $\zeta=0.1$ is adopted.

The second-order Fermi acceleration is limited by various processes. 
The proton-proton relaxation and Coulomb scattering, 
as seen in equations~(\ref{eq:trelax}) and (\ref{eq:tcoul}), 
are also possible processes to suppress such a stochastic acceleration.
If the radiation energy is high enough to damp the magnetic turbulent waves by 
the Compton scattering, the Compton drag can prevent the protons from accelerating 
\citep{1998PhRvD..57.3219T}. We approximate the timescale for the Compton drag as
\begin{equation}
t_{\rm Cd}=\frac{B^2}{(4/3)\sigma_Tn_pU_{\gamma}c}\max\left(1,\frac{1}{\tau_{p}}\right),
%=\frac{4\pi}{3}\frac{1}{\eta_{}}\left(\frac{v_{\rm K}}{c}\right)^2\left(\frac{r}{c}\right),
\label{eq:tcd}
\end{equation}
where $U_\gamma\sim{L}t_{\rm inf}/(4\pi{r_{\rm p}^3}/3)$, with luminosity $L=\eta\dot{M}c^2$, 
is the radiation energy density. Note that $\eta$ is the mass-to-energy conversion efficiency, 
which takes a different value for each phase.

The protons potentially escape from the acceleration region via spatial diffusion.   
For isotropic turbulence, the diffusion time of the protons is given by  \citep{2015ApJ...806..159K},
\begin{equation}
t_{\rm diff}=9\zeta\frac{r}{c}\left(\frac{r_{\rm L}}{r}\right)^{s-2}\gamma^{s-2}.
\label{eq:tdiff}
\end{equation}
The proton synchrotron emission and inelastic $pp$ processes are adopted 
as a promising cooling mechanism in the TDR. 
Respective cooling timescales are given by 
\begin{eqnarray}
t_{\rm sync}&=& \frac{3}{4}\left(\frac{m_{p}}{m_{e}}\right)^{3}\frac{m_{e}c^{2}}{c\sigma_{\rm T}U_{\rm B}}
\frac{1}{\gamma}
\label{eq:tsync}
\end{eqnarray}
and
\begin{eqnarray}
t_{pp}&=&\frac{1}{n_{p}\sigma_{pp}cK_{pp}},
\label{eq:tpp}
\end{eqnarray}
where $U_{\rm B}=B^{2}/(8\pi)$ and $K_{pp}\sim0.5$ 
are the energy density of the magnetic fields and the proton inelasticity, respectively, 
and the total cross section of the $pp$ process is represented by 
$\sigma_{pp}
\simeq
10^{-27}\,{\rm cm^2}\,
[34.3+
1.88
\log
(E_{p}/1\,{\rm TeV})
+ 
0.25\log^{2}
(
E_p/1\,{\rm TeV}
)
]
[
1-(
E_{pp,{\rm thr}}/E_{p}
)^{4}
]^{2}
$
for $E_{p}\ge{E_{pp,{\rm thr}}}$. Here $E_p=\gamma{m_p}c^2$ 
is the proton energy and $E_{pp,{\rm thr}}=1.22\,{\rm GeV}$ \citep{2006PhRvD..74c4018K}. 
The $p\gamma$ cooling timescale is given by \citep{1996ApJ...456..106D}
\begin{equation}
t_{p\gamma}=\left[\frac{c}{2\gamma^2}\bar{\epsilon}_{\rm pk}\Delta\bar{\epsilon}_{\rm pk}\sigma_{\rm pk}K_{\rm pk}\int_{\bar{\epsilon}_{\rm pk}/(2\gamma)}^{\infty}dE_\gamma\frac{N_\gamma(E_\gamma)}{E^2_\gamma}\right]^{-1},
\label{eq:tpgamma}
\end{equation}
where $\bar{\epsilon}_{\rm pk}\sim0.3\,{\rm GeV}$, $\sigma_{\rm pk}\sim5\times10^{-28}\,{\rm cm^2}$, $K_{\rm pk}\sim0.2$, and $\Delta\bar{\epsilon}_{\rm pk}\sim0.2\,{\rm GeV}$. For the highly optically thin region during the RIAF phase ($\tau_{p}={n}_p\sigma_{\rm T}r\sim3.1\times10^{-2}\,(\alpha/0.1)^{-1}(m_*/M_\odot)^{1/6}(r_*/R_\odot)^{-1/2}(r/r_{\rm p})^{-1/2}(\beta/1)^{1/2}(\dot{m}/0.01)(M_{\rm bh}/10^7M_\odot)^{1/3}$), the photon-proton interaction is by definition inefficient. Therefore, $t_{\rm p\gamma}$ is much longer than the timescales by the other interactions. We neglect the effect of cooling by the photon-proton interaction for the RIAF phase (see \citealt{2015ApJ...806..159K} for the energy dependence of $t_{\rm p\gamma}$ for the RIAFs in the low-luminosity AGNs).

By using equations~(\ref{eq:taccl}) and (\ref{eq:tpp}), 
we obtain the Lorenz factor of the proton at $t_{\rm accl}=t_{pp}$:
\begin{eqnarray}
\gamma_{pp}(r)
&=&
\left(\frac{\zeta}{K_{pp}}\right)^{1/(2-s)}
\left(\frac{v_{\rm A}}{c}\right)^{6}
\left(\frac{r}{r_{\rm L}}\right)
%\left(\frac{r_{\rm t}}{r_{\rm S}}\right)^{3}
\left(\frac{1}{n_{p}\sigma_{pp}r}\right)^{1/(2-s)}.
%\left(\frac{r}{r_{\rm t}}\right)^{(1-s)/{(2-s)}}.
\label{eq:gammapp}
\end{eqnarray}
This is estimated at $r=r_{\rm c}$ to be
\begin{eqnarray}
\gamma_{pp}(r_{\rm c})
&\sim&
1.2\times10^{-11}
\left(\frac{\zeta}{0.1}\right)^{3}
\left(\frac{\mathcal{B}}{3}\right)^{-7/2}
\left(\frac{\beta}{1.0}\right)^{33/2}
\left(\frac{\dot{m}}{\dot{m}_{\rm fb}}\right)^{-5/2} \nonumber \\
&\times&
\left(\frac{M_{\rm bh}}{10^7\,M_\odot}\right)^{73/12} 
\left(\frac{m_*}{M_\odot}\right)^{41/12} 
\left(\frac{r_*}{R_\odot}\right)^{-31/4},
%\left(\frac{r_{\rm t}}{r_{\rm S}}\right)^{(5s-4)/{(8-4s)}}\left(\frac{r}{r_{\rm t}}\right)^{-(s+2)/{(4-2s)}}.
\end{eqnarray}
where we adopt $s=5/3$ and $\sigma_{pp}\approx3.6\times10^{-26}\,{\rm cm^2}$. 
Hence $\gamma_{pp}(r_{\rm c})$ cannot be larger than unity because of very efficient 
proton-proton cooling. Figure~\ref{fig:timescale} shows the dependence of the characteristic 
cooling timescales normalized by $t_{\rm accl}$ on the proton energy. Each panel 
shows the normalized timescales of possible sites to produce the high-energy particles. 
Panel (a) depicts the normalized timescales of the first shock of the debris circularization phase of TDEs. 
The details of the other panels are described in the later corresponding sections. We note from 
panel (a) that the first (strongest) shock during the debris circularization is unlikely to produce 
the protons and neutrinos in the reasonable energy range, since the proton-proton collision cooling 
time is much shorter than the acceleration time.

%
%%%%%%%%%%%%%%%%%%%%%%%%%%%
\subsection{Super-Eddington accretion phase}
%%%%%%%%%%%%%%%%%%%%%%%%%%%
%
In an optically and geometrically thick accretion flow with the mass accretion rate 
exceeding the Eddington limit, the photons are trapped and restored as an entropy 
in the accreting gas without being radiated away. In other words, the advective cooling 
dominates the radiative cooling. The photon trapping radius is given by equating the 
radiative diffusion timescale with the accretion timescale as
\begin{eqnarray}
\frac{r_{\rm trap}}{r_{\rm S}}=\frac{3}{2}\dot{m}\left(\frac{H}{r}\right)
\sim64\,
\left(\frac{H/r}{1.0}\right)
\left(\frac{\dot{m}}{\dot{m}_{\rm fb}}\right)
\left(\frac{M_{\rm bh}}{10^7\,M_\odot}\right)^{-3/2}
\left(\frac{m_*}{M_\odot}\right)^{2}\left(\frac{r_*}{R_\odot}\right)^{-3/2}
\nonumber 
\label{eq:rtrap}
\end{eqnarray}
with equation~(\ref{eq:mdotfb}).
As far as $\dot{m}\gg1$, one can see $r_{\rm trap}\gg{r_{\rm t}}$ for $H/r\sim1$ 
so that the TDE disk should be the super-Eddington accretion flow.

In the TDE context, we estimate the number density of the super-Eddington disk as
\begin{eqnarray}
n_{p}&=&\frac{\dot{M}}{2\pi{r^2}v_rm_{p}} 
\sim2.7\times10^{13}\,{\rm cm^{-3}}
\nonumber \\
&\times&
\left(\frac{\alpha}{0.1}\right)^{-1}
\left(\frac{\beta}{1.0}\right)^{3/2} 
\left(\frac{\dot{m}}{\dot{m}_{\rm fb}}\right) 
\left(\frac{M_{\rm bh}}{10^7\,M_\odot}\right)^{-3/2} 
\left(\frac{m_*}{M_\odot}\right)^{5/2} 
\left(\frac{r_*}{R_\odot}\right)^{-3}
\left(\frac{r}{r_{\rm p}}\right)^{-3/2}
\label{eq:seafnp}
\end{eqnarray}
by using equation (\ref{eq:mdotfb}), where $v_r=\alpha\,v_{\rm K}(r)$ is the radial drift velocity, 
which corresponds to that of the simplest solution for the slim-disk model 
\citep{1988ApJ...332..646A,1999ApJ...516..420W,2006ApJ...648..523W}.
From equation (\ref{eq:gammapp}), the Lorenz factor $\gamma_{pp}$ 
at the pericenter radius $r=r_{\rm p}$ is estimated to be
\begin{eqnarray}
\gamma_{pp}(r_{\rm p})
&\sim&
1.6\times10^{}\,
\left(\frac{\alpha}{0.1}\right)^{3}
\left(\frac{\zeta}{0.1}\right)^{3}
\left(\frac{\mathcal{B}}{3}\right)^{-7/2}
\left(\frac{\beta}{1.0}\right)^{7/4} \nonumber \\
&\times&
\left(\frac{\dot{m}}{\dot{m}_{\rm fb}}\right)^{-5/2} 
\left(\frac{M_{\rm bh}}{10^7\,M_\odot}\right)^{65/12} 
\left(\frac{m_*}{M_\odot}\right)^{-53/12} 
\left(\frac{r_*}{R_\odot}\right)^{2/3},
%\left(\frac{r_{\rm t}}{r_{\rm S}}\right)^{(5s-4)/{(8-4s)}}\left(\frac{r}{r_{\rm t}}\right)^{-(s+2)/{(4-2s)}}.
\label{eq:gammapp2}
\end{eqnarray}
where $s=5/3$ and $\sigma_{pp}\approx3.6\times10^{-26}\,{\rm cm^2}$ are adopted. 
Taking account of the equipartition assumption of the magnetic field (see  equation~\ref{eq:bfield}), we find 
$\gamma_{pp}(r)\propto{B}^{7}\propto{T^{7/2}}$ together with 
$T\sim10^{11}\,{\rm K}$ with $M_{\rm bh}=10^7\,M_{\odot}$ and $\mathcal{B}=3$. 
However, the disk temperature of the super-Eddington accretion flow increases up to $\sim10^{8-9}\,{\rm K}$ 
by the latest three-dimensional radiation magneto-hydrodynamic (MHD) simulations of super-Eddington accretion flow around the SMBH with $M_{\rm bh}=5.0\times10^8\,M_\odot$ \citep{2017arXiv170902845J}. Since the disk temperature becomes 
lower as the black hole mass becomes lower, $\gamma_{pp}(r_{\rm p})$ cannot be larger than unity for $M_{\rm bh}\lesssim10^8\,M_\odot$. 
This suggests that the cooling by the proton-proton collision is very efficient in the super-Eddington 
accretion flow. Panel (b) of Figure~\ref{fig:timescale} depicts the dependence of characteristic 
timescales normalized by $t_{\rm accl}$ on the proton energy in the super-Eddington accretion flow. 
From the figure, the super-Eddington accretion flows are unlikely to accelerate the protons to the 
ultrarelativistic regime and thus produce the neutrinos in the reasonable energy range.

%
%%%%%%%%%%%%%%%%%%%%%%%%%%%%
\subsection{Radiatively inefficient accretion flow phase}
%%%%%%%%%%%%%%%%%%%%%%%%%%%%
%

The RIAFs are very hot and optically thin so that they 
can produce high-energy emissions. 
This is because a heat produced via turbulent viscosity is stored as 
entropy and transported inwardly with accretion. The original model 
of the RIAF is a one-dimensional, optically thin, advection-dominated 
accretion flow (ADAF) \citep{1994ApJ...428L..13N,1995ApJ...452..710N}. 
The number density of the RIAFs is estimated to be 
\begin{eqnarray}
n_{p}&=&\frac{\dot{M}}{2\pi\alpha{r^2}v_{\rm K}m_{p}}
\sim3.2\times10^{9}\,{\rm cm^{-3}}
\nonumber \\
&\times&
\left(\frac{\alpha}{0.1}\right)^{-1}
\left(\frac{\beta}{1.0}\right)^{3/2} 
\left(\frac{\dot{m}}{0.01}\right) 
\left(\frac{m_*}{M_\odot}\right)^{1/2} 
\left(\frac{r_*}{R_\odot}\right)^{-3/2}
\left(\frac{r}{r_{\rm p}}\right)^{3/2},
\label{eq:riafnp}
\end{eqnarray}
where the radial velocity is assumed to be $\alpha\,{v}_{\rm K}$ as 
a simplest solution of the ADAF.

Substituting equations (\ref{eq:rl}), (\ref{eq:bfield}), (\ref{eq:alfven}), 
and (\ref{eq:riafnp}) into equation (\ref{eq:gammapp}), the Lorenz factor at 
$t_{pp}=t_{\rm accel}$ during the RIAF phase is estimated to be
\begin{eqnarray}
\gamma_{pp}(r_{\rm p})
&\sim&
1.1\times10^{11}\,
\left(\frac{\zeta}{0.1}\right)^{3}
\left(\frac{\mathcal{B}}{3}\right)^{-7/2}
\left(\frac{\alpha}{0.1}\right)^{5/2}
\left(\frac{\beta}{1.0}\right)^{7/4}
\left(\frac{\dot{m}}{0.01}\right)^{-5/2}
\nonumber \\
&\times&
\left(\frac{M_{\rm bh}}{10^7\,M_\odot}\right)^{5/3}
\left(\frac{m_{*}}{M_\odot}\right)^{7/12}
\left(\frac{r_{*}}{R_\odot}\right)^{-7/4},
\label{eq:ppeq}
\end{eqnarray}
where $s=5/3$ and $\sigma_{pp}\approx3.6\times10^{-26}\,{\rm cm^2}$ are adopted. 
By using equations (\ref{eq:taccl}) and (\ref{eq:tdiff}), we obtain the Lorenz factor 
at $t_{\rm diff}=t_{\rm accel}$:
\begin{eqnarray}
\gamma_{\rm diff}(r)
=
\left(3\zeta\frac{{v}_{\rm A}}{c}\right)^{1/(2-s)}\left(\frac{r}{r_{\rm L}}\right).
\end{eqnarray}
Substituting equations (\ref{eq:rl}), (\ref{eq:bfield}), (\ref{eq:alfven}), and (\ref{eq:riafnp}) into 
equations (\ref{eq:taccl}) and (\ref{eq:tdiff}), this is estimated at $r=r_{\rm p}$ to be
\begin{eqnarray}
\gamma_{\rm diff}(r_{\rm p})
&\sim&
4.8\times10^5\,
\left(\frac{\alpha}{0.1}\right)^{-1/2}
\left(\frac{\zeta}{0.1}\right)^{3}
\left(\frac{\mathcal{B}}{3}\right)^{-2}
\left(\frac{\dot{m}}{0.01}\right)^{1/2}
\left(\frac{\beta}{1.0}\right)^{7/4}
\nonumber \\
&\times&
\left(\frac{M_{\rm bh}}{10^7\,M_\odot}\right)^{5/3}
\left(\frac{r}{r_{\rm p}}\right)^{-7/4}
\left(\frac{m_{*}}{M_\odot}\right)^{7/12}
\left(\frac{r_{*}}{R_\odot}\right)^{-7/4},
\label{eq:diffeq}
\end{eqnarray}
where $s=5/3$ and $\sigma_{pp}\approx3.6\times10^{-26}\,{\rm cm^2}$ are adopted. 
Panel (d) of Figure~\ref{fig:timescale} depicts the dependence of characteristic 
timescales normalized by $t_{\rm accl}$ on the proton energy in the RIAF. 
We note from the figure that the diffusion timescale is the shortest among the 
timescales of the other mechanisms, which prevents the protons from accelerating. 
In this case, the protons can be accelerated up to 
$E_{p,\rm {diff}}=\gamma_{\rm diff}(r_{\rm p})\,m_pc^2\simeq0.45\,{\rm PeV}$.

%
%%%%%%%%%%%%%%%%%%%%
\subsection{Magnetically Arrested Disks}
\label{sec:mad}
%%%%%%%%%%%%%%%%%%%%
%
%A significant amount of poloidal magnetic flux around a black hole 
A large-scale poloidal magnetic field prevents gas from 
accreting continuously at a magnetospheric radius, which is far 
outside the event horizon of the black hole \citep{1974ApSS..28...45B}. 
Around the magnetospheric radius, the gas flow breaks up into a blob-like 
stream and moves inward by diffusing via magnetic interchanges through the magnetic field. 
\citet{2003PASJ...55L..69N} called such a disrupted accretion flow a magnetically arrested disk (MAD). 
The MAD state has been tested by numerical MHD simulations 
\citep{2012MNRAS.423.3083M,2015MNRAS.454L...6M,2018MNRAS.478.1837M}.
 %as a result of the cumulative action of the accretion flow, 
 %and that the magnetic field is dynamically dominant. 
 % The field is prevented from escaping by the continued 
 % inward pressure of accretion. We call such a disrupted accretion flow a “magnetically arrested disk” or MAD for short.

The main difference from the previous two cases is how to estimate the strength of the magnetic field. 
By equating the gravitational force per unit area of the radially accreting mass 
$GM\rho_{p,{\rm MAD}}H/r^2$ with the magnetic energy density $B^2/(8\pi)$, 
the square of the magnetic field strength of the MAD state is then given by
\begin{eqnarray}
B^2_{\rm MAD}
=
2\sqrt{2}\pi
\left(\frac{\alpha}{\epsilon}\right)
\left(\frac{H}{r}\right)
\rho_{p}v^2_{\rm ff}(r),
%4\pi\left(\frac{1}{\epsilon}\right)\left(\frac{H}{r}\right)\rho_pv^2_{\rm ff},
%\\B^2&=&\frac{4\pi}{3\mathcal{B}}\rho_{p}v_{\rm ff}(r)^2,
\label{eq:bmad}
\end{eqnarray}
where $v_{\rm ff}(r)=\sqrt{2GM_{\rm bh}/r}$ is the freefall velocity and the mass conservation law gives the local density estimated at the magnetosphere as $\rho_{p,{\rm MAD}}=(v_{r}/v_{r,{\rm MAD}})\rho_p$ with the radial magnetic diffusion velocity $v_{r,{\rm MAD}}=\epsilon\,{v}_{\rm ff}(r)$ with $\epsilon\lesssim0.01$ \citep{2003PASJ...55L..69N}. Comparing with equation~(\ref{eq:bfield}), we obtain $B^2_{\rm MAD}/B^2=(3/\sqrt{2})\mathcal{B}(\alpha/\epsilon)(H/r)$. Since $H/r\sim1$ for the super-Eddington accretion flows and RIAFs, the field strength in the MAD state is $\sim\sqrt{\alpha/\epsilon}$ times larger than that of the dipole field for the given plasma beta.

%
%%%%%%%%%%%%%%%%%%%%%%%%%%%%%%%%%
\subsubsection{Super-Eddington Magnetically Arrested Disks}
\label{sec:semad}
%%%%%%%%%%%%%%%%%%%%%%%%%%%%%%%%%
%
Here we apply the MAD state for the super-Eddington accretion flow. 
The number density of the super-Eddington MAD is given by
\begin{eqnarray}
n_{p,{\rm MAD}}=\frac{\dot{M}}{2\pi{r^2}v_{r,{\rm MAD}}m_{p}} 
&\sim&
1.9\times10^{14}\,{\rm cm^{-3}}
\left(\frac{\epsilon}{0.01}\right)^{-1}
\left(\frac{M_{\rm bh}}{10^7\,M_\odot}\right)^{-3/2}
\nonumber \\
&\times&
 \left(\frac{\beta}{1.0}\right)^{3/2} 
\left(\frac{\dot{m}}{\dot{m}_{\rm fb}}\right) 
\left(\frac{m_*}{M_\odot}\right)^{5/2} 
\left(\frac{r_*}{R_\odot}\right)^{-3}
\left(\frac{r}{r_{\rm p}}\right)^{-3/2}
\label{eq:seafmadnp}
\end{eqnarray}
by using equation (\ref{eq:mdotfb}). 
By substituting equation~(\ref{eq:seafmadnp}) into equation (\ref{eq:bmad}), 
we estimate $B_{\rm MAD}$ at $r=r_{\rm p}$ for $H/r\sim1$ as
\begin{eqnarray}
B_{\rm MAD}(r_{\rm p})
&\sim&
2.2\times10^6\,{\rm Gauss}\,
\left(\frac{\epsilon}{0.01}\right)^{-1/2}
\left(\frac{M_{\rm bh}}{10^7\,M_\odot}\right)^{-5/12}
\nonumber \\
&\times&
\left(
\frac{\beta}{1.0}
\right)^{5/4}
\left(\frac{\dot{m}}{\dot{m}_{\rm fb}}\right)^{1/2}
\left(\frac{m_{*}}{M_\odot}\right)^{17/12}
\left(\frac{r_{*}}{R_\odot}\right)^{-2}.
\label{eq:bmad2}
\end{eqnarray}
Panel (e) of Figure~\ref{fig:timescale} depicts the dependence 
of characteristic timescales normalized by $t_{\rm accl}$ 
on the proton energy in the super-Eddington MAD state. We note 
from the figure that the Compton drag is the most efficient mechanism 
to prevent the protons from accelerating. By using equations~(\ref{eq:taccl}) 
and (\ref{eq:tcd}), we obtain the Lorenz factor of the proton 
at $t_{\rm accl}=t_{\rm Cd}$:
\begin{eqnarray}
\gamma_{\rm Cd}(r)
&=&
\left(
\pi
\frac{\epsilon}{\eta}
\frac{\zeta}{\tau_p}
\right)^{1/(2-s)}
\left(\frac{v_{\rm A}}{c}\right)^{2/(2-s)}
\left(\frac{r}{r_{\rm L}}\right)
\left(\frac{B^2r^3}{\dot{M}c^2}\frac{v_{\rm ff}}{r}\right)^{1/(2-s)}.
\label{eq:gammasync}
\end{eqnarray}
It is estimated at $r=r_{\rm p}$ to be
\begin{eqnarray}
\gamma_{\rm Cd}(r_{\rm p})
&\sim&
2.6\,
\left(\frac{\epsilon}{0.01}\right)^{5/2}
\left(\frac{\eta_{\rm MAD}}{0.15}\right)^{-3}
\left(\frac{\zeta}{0.1}\right)^{3}
\left(\frac{\beta}{1.0}\right)^{19/4} 
\left(\frac{M_{\rm bh}}{10^7\,M_\odot}\right)^{89/12} 
\nonumber \\
&\times&
\left(\frac{\dot{m}}{\dot{m}_{\rm fb}}\right)^{-5/2} 
\left(\frac{m_*}{M_\odot}\right)^{-41/12} 
\left(\frac{r_*}{R_\odot}\right)^{-1},
\label{eq:gammacdrag}
\end{eqnarray}
where we adopt $s=5/3$ as the spectral index and 
$\eta_{\rm MAD}=0.15$ as the radiative efficiency, 
which is obtained from three-dimensional general 
relativistic radiation MHD simulations \citep{2015MNRAS.454L...6M}. 
The second efficient mechanism is the proton-proton cooling. 
From equation (\ref{eq:gammapp}) with equation (\ref{eq:bmad}), $\gamma_{pp}$ 
is estimated at $r=r_{\rm p}$ to be
\begin{eqnarray}
\gamma_{pp}(r_{\rm p})
&\sim&
2.3\times10^2\,
\left(\frac{\epsilon}{0.01}\right)^{5/2}
\left(\frac{\dot{m}}{\dot{m}_{\rm fb}}\right)^{-5/2}
\left(\frac{M_{\rm bh}}{10^7\,M_\odot}\right)^{65/12} 
\nonumber \\
&\times&
\left(\frac{\zeta}{0.1}\right)^{3}
\left(\frac{\beta}{1.0}\right)^{7/4} 
\left(\frac{m_*}{M_\odot}\right)^{-53/12} 
\left(\frac{r_*}{R_\odot}\right)^{2},
\label{eq:gammappmad}
\end{eqnarray}
where $s=5/3$ and $\sigma_{pp}\approx3.6\times10^{-26}\,{\rm cm^2}$ are adopted. 
The third efficient mechanism is the synchrotron cooling. By using equations~(\ref{eq:taccl}) and (\ref{eq:tsync}), 
we obtain the Lorenz factor of the proton at $t_{\rm accl}=t_{\rm sync}$:
\begin{eqnarray}
\gamma_{\rm sync}(r)
&=&
\left(
\frac{3}{4}\zeta
\right)^{1/(3-s)}
\left(
\frac{m_{p}}{m_{e}}
\right)^{2/(3-s)}
\left(\frac{{m_{p}}c^2}{r\sigma_{\rm T}U_{\rm B}}\right)^{1/(3-s)}
\left(\frac{v_{\rm A}}{c}\right)^{2/(3-s)}
\left(\frac{r}{r_{\rm L}}\right)^{(2-s)/(3-s)}.
\label{eq:gammasync}
\end{eqnarray}
It is estimated at $r=r_{\rm p}$ to be
\begin{eqnarray}
\gamma_{\rm sync}(r_{\rm p})
&\sim&
9.4\times10^{4}\,
\left(\frac{\epsilon}{0.01}\right)^{5/8}
\left(\frac{\zeta}{0.1}\right)^{3/4}
\left(\frac{\beta}{1.0}\right)^{-5/16} 
\left(\frac{M_{\rm bh}}{10^7\,M_\odot}\right)^{41/48} 
\nonumber \\
&\times&
\left(\frac{\dot{m}}{\dot{m}_{\rm fb}}\right)^{-5/8} 
\left(\frac{m_*}{M_\odot}\right)^{-65/48} 
\left(\frac{r_*}{R_\odot}\right)^{5/4},
\label{eq:gammasync2}
\end{eqnarray}
where $s=5/3$ is adopted.

It is clear from equation (\ref{eq:gammacdrag}) that the protons can be 
accelerated to at most $\sim2.6\,{\rm GeV}$ if $M_{\rm  bh}\le10^7\,M_\odot$. 
We find from the above three equations that $\gamma_{{\rm Cd}}$ and 
$\gamma_{pp}$ rapidly increase with the black hole mass, whereas 
$\gamma_{\rm sync}$ weakly depends on it. We also find that 
$\gamma_{\rm Cd}\lesssim\gamma_{pp}$ for $M_{\rm bh}\lesssim10^8\,M_\odot$, 
meaning the Compton drag is more efficient than the proton-proton cooling 
in the given mass range. 
Next we compare $\gamma_{\rm Cd}$ with $\gamma_{\rm sync}$. 
The synchrotron cooling timescale is shorter than the Compton drag timescale 
if $\gamma_{\rm Cd}/\gamma_{\rm sync}>1$.
Because we find from equations~(\ref{eq:gammacdrag}) and (\ref{eq:gammasync2}) that 
$\gamma_{\rm Cd}/\gamma_{\rm sync}\propto{M_{\rm bh}^{105/16}}$, there is a critical 
value of black hole mass, $M_{\rm bh}\sim10^{7.7}\,M_{\odot}$, where $\gamma_{\rm Cd}/\gamma_{\rm sync}=1$. 
Therefore, the Compton drag is more efficient than the synchrotron cooling if $M_{\rm bh}<10^{7.7}\,M_{\odot}$. 
In this case, the protons can be accelerated to 
$E_{p,{\rm Cd}}=\gamma_{\rm Cd}(r_{\rm p})\,m_pc^2\simeq2.2\,{\rm TeV}(M_{\rm bh}/10^{7.4}\,M_\odot)^{89/12}$. 
On the other hand, the synchrotron cooling is more efficient if $M_{\rm bh}\gtrsim10^{7.7}\,M_{\odot}$.
In this case, the protons can be accelerated up to 
$E_{p,\rm {sync}}=\gamma_{\rm sync}(r_{\rm p})\,m_pc^2\simeq0.35\,{\rm PeV}(M_{\rm bh}/10^{7.7}\,M_\odot)^{41/48}$.

%
%%%%%%%%%%%%%%%%%%%%%%%%%%%%%%%%%%
\subsubsection{Radiatively Inefficient Magnetically Arrested Disks}
%%%%%%%%%%%%%%%%%%%%%%%%%%%%%%%%%
%
Next we consider the RIAF with the MAD state. 
We call it "radiatively inefficient MAD" in what follows. 
The number density of the radiatively inefficient MAD is given by
\begin{eqnarray}
n_{p,{\rm MAD}}&=&\frac{\dot{M}}{2\pi{r^2}v_{r,{\rm MAD}}m_{p}} 
\sim2.2\times10^{10}\,{\rm cm^{-3}}\,
\left(\frac{\epsilon}{0.01}\right)^{-1}
\left(\frac{\dot{m}}{0.01}\right)
\left(\frac{r}{r_{\rm p}}\right)^{-3/2} 
\nonumber \\
&\times&
%\left(\frac{M_{\rm bh}}{10^7\,M_\odot}\right)^{-3/2} 
\left(\frac{\beta}{1.0}\right)^{3/2} 
\left(\frac{m_*}{M_\odot}\right)^{1/2} 
\left(\frac{r_*}{R_\odot}\right)^{-3/2},
\label{eq:riafmadnp}
\end{eqnarray}
where equation (\ref{eq:mmdot}) is used for $\dot{M}$.
We confirm that the radiatively inefficient MAD is optically thin because of 
$\tau_{p}=n_{p,\rm{MAD}}\sigma_{\rm T}r\sim2.2\times10^{-1}\,(\epsilon/0.01)^{-1}(\dot{m}/0.01)(M_{\rm bh}/10^7M_\odot)^{1/3}
(r/r_{\rm p})^{-1/2}(\beta/1.0)^{1/2}(m_*/M_\odot)^{1/6}(r_*/R_\odot)^{-1/2}$. 
By substituting equation~(\ref{eq:riafmadnp}) into equation (\ref{eq:bmad}), 
we estimate the magnetic field strength at $r=r_{\rm p}$ for $H/r\sim1$ as
\begin{eqnarray}
B_{\rm MAD}(r_{\rm p})
&\sim&
9.1\times10^3\,{\rm G}\,
\left(\frac{\epsilon}{0.01}\right)^{-1/2}
\left(\frac{\alpha}{0.1}\right)^{1/2}
\left(
\frac{\beta}{1.0}
\right)^{5/4}
\left(\frac{\dot{m}}{0.01}\right)^{1/2}
\left(\frac{M_{\rm bh}}{10^7\,M_\odot}\right)^{1/3}
\nonumber \\
&\times&
\left(\frac{m_{*}}{M_\odot}\right)^{5/12}
\left(\frac{r_{*}}{R_\odot}\right)^{-5/4}
\label{eq:bmadriaf}
\end{eqnarray}

\begin{figure}[!ht]
\centering
\includegraphics[width=7.1cm]{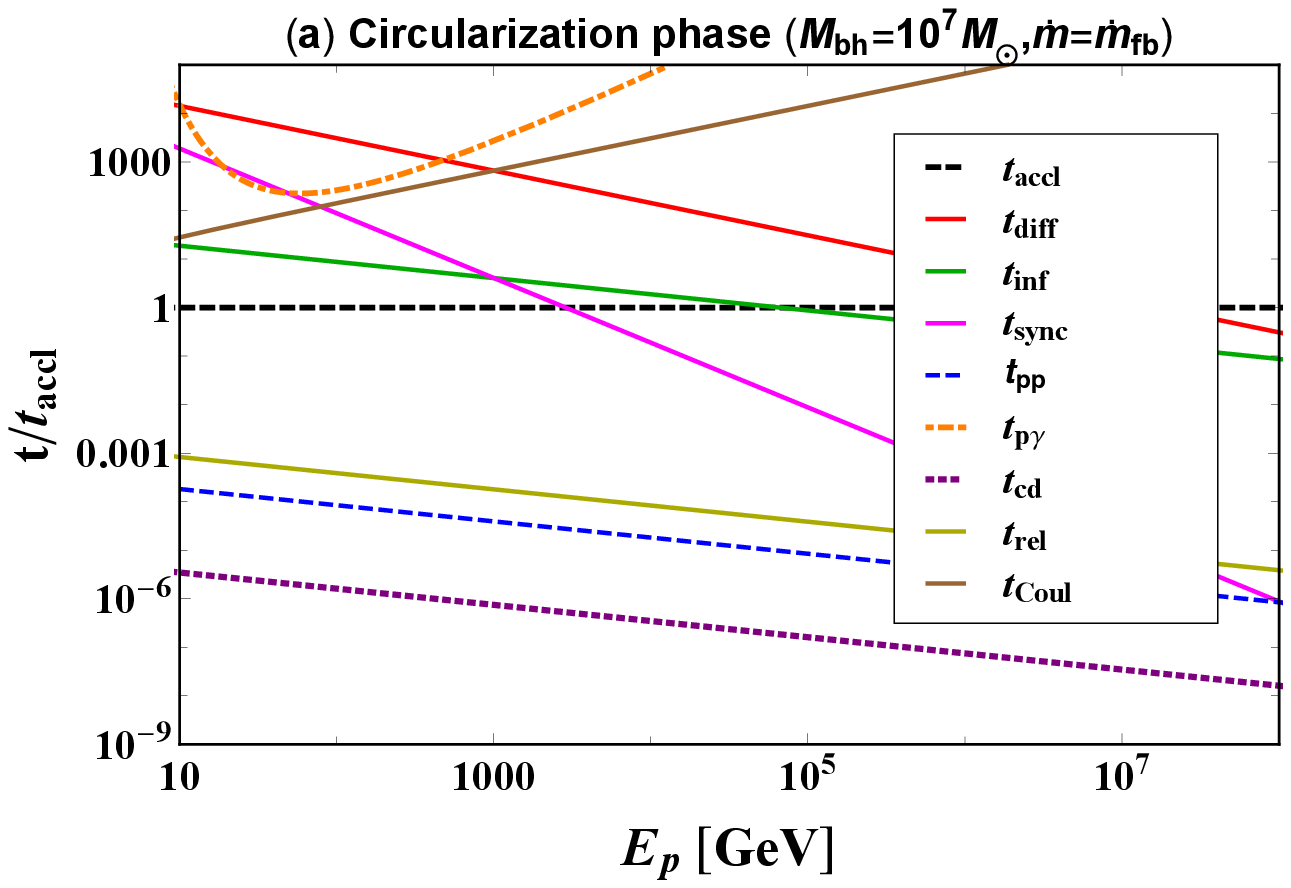}
\includegraphics[width=7.1cm]{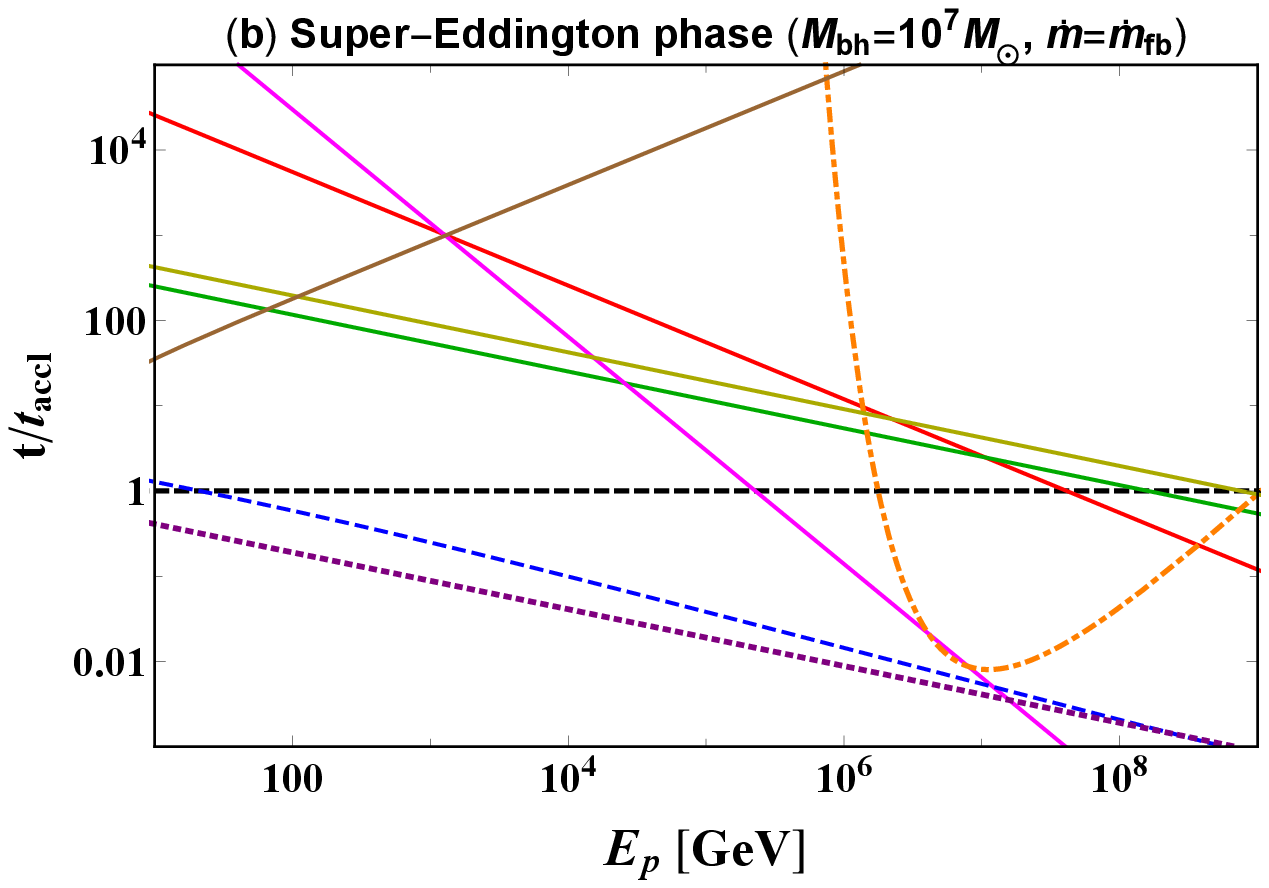}
\\
\includegraphics[width=7.1cm]{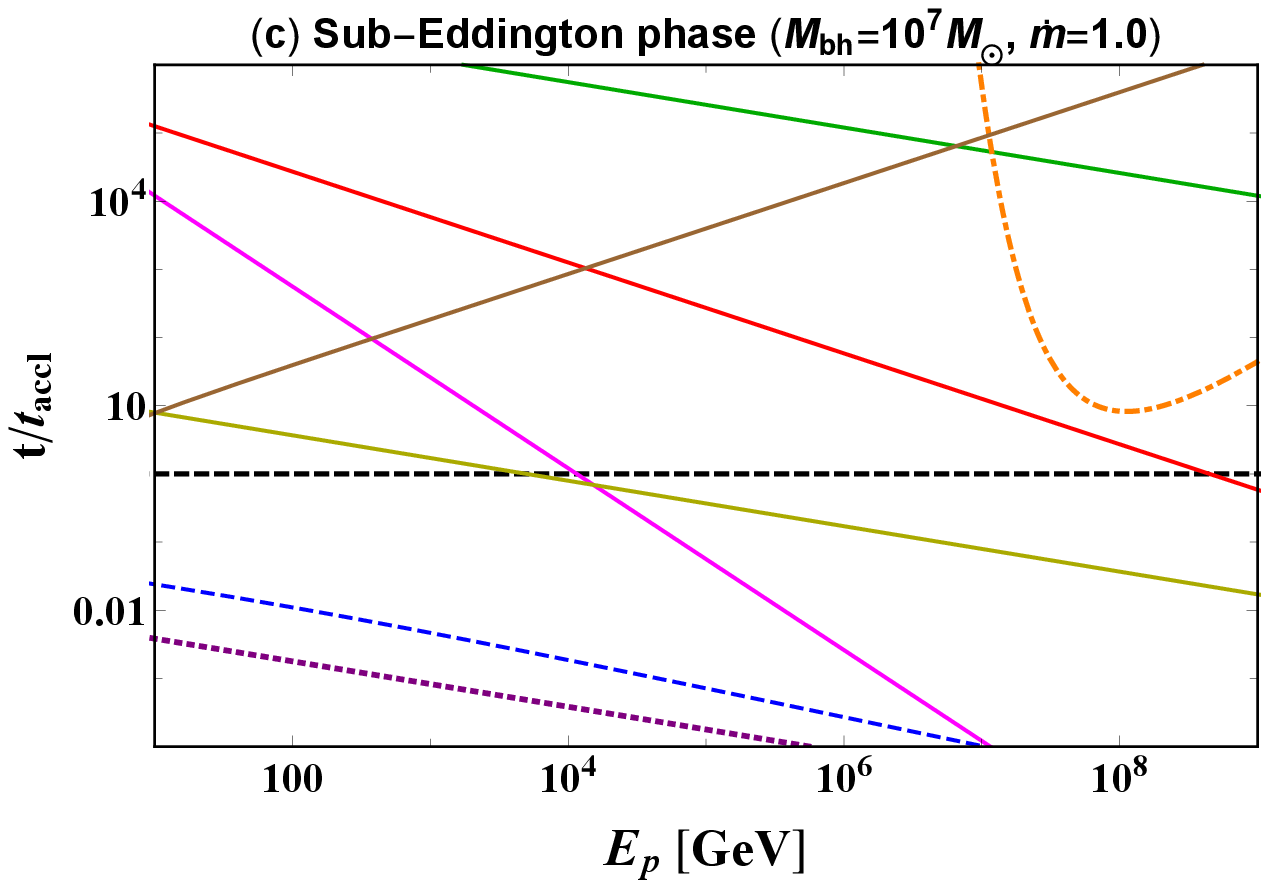}
\includegraphics[width=7.1cm]{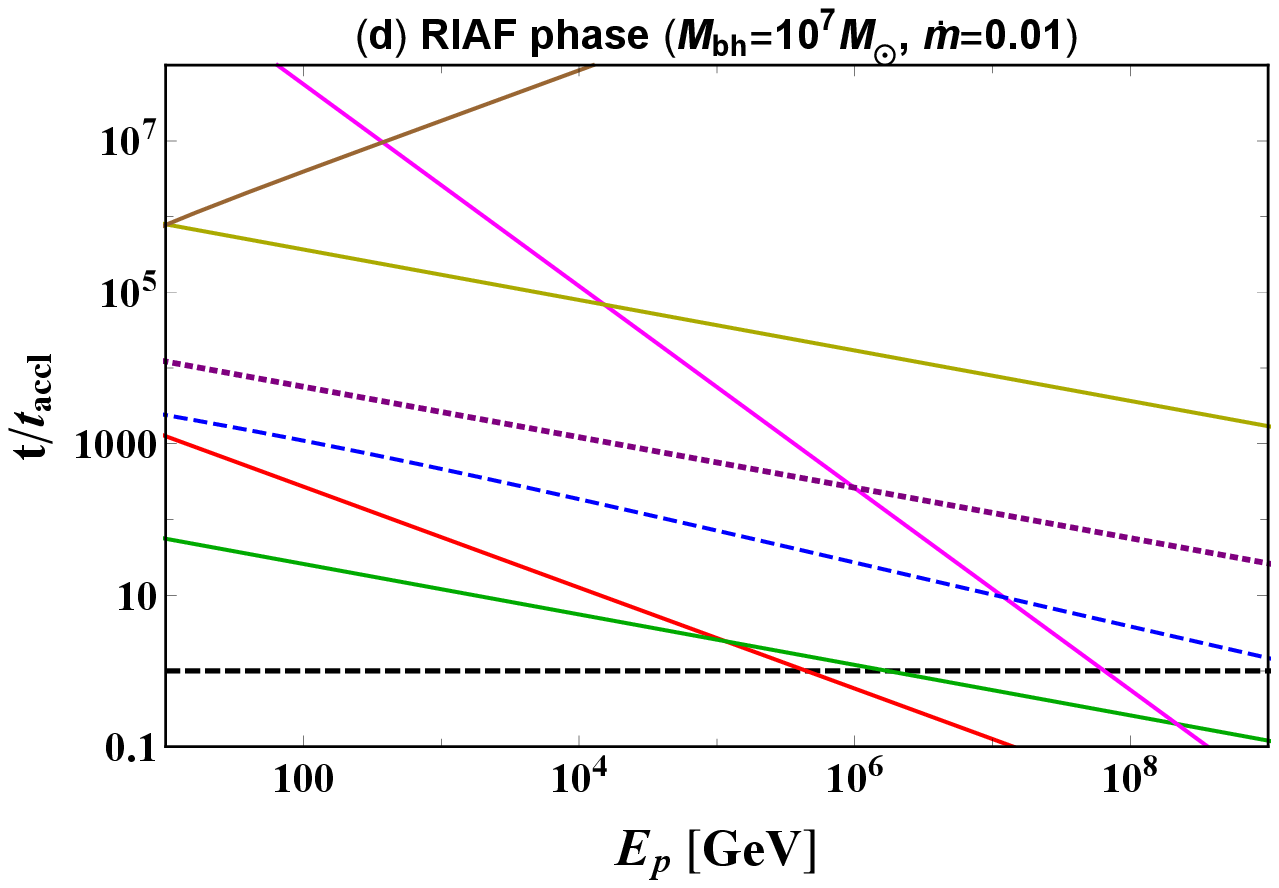}
\\
\includegraphics[width=7.1cm]{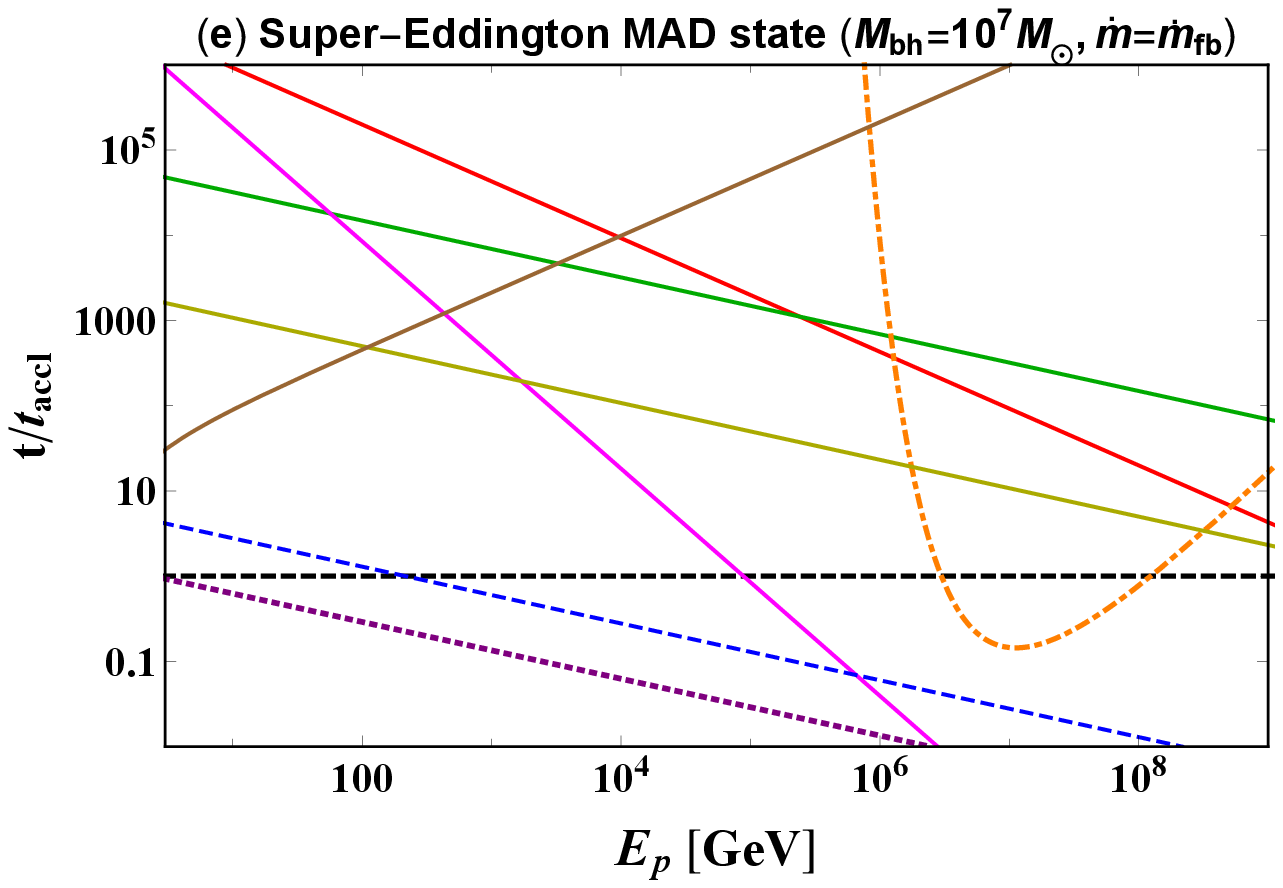}
\includegraphics[width=7.1cm]{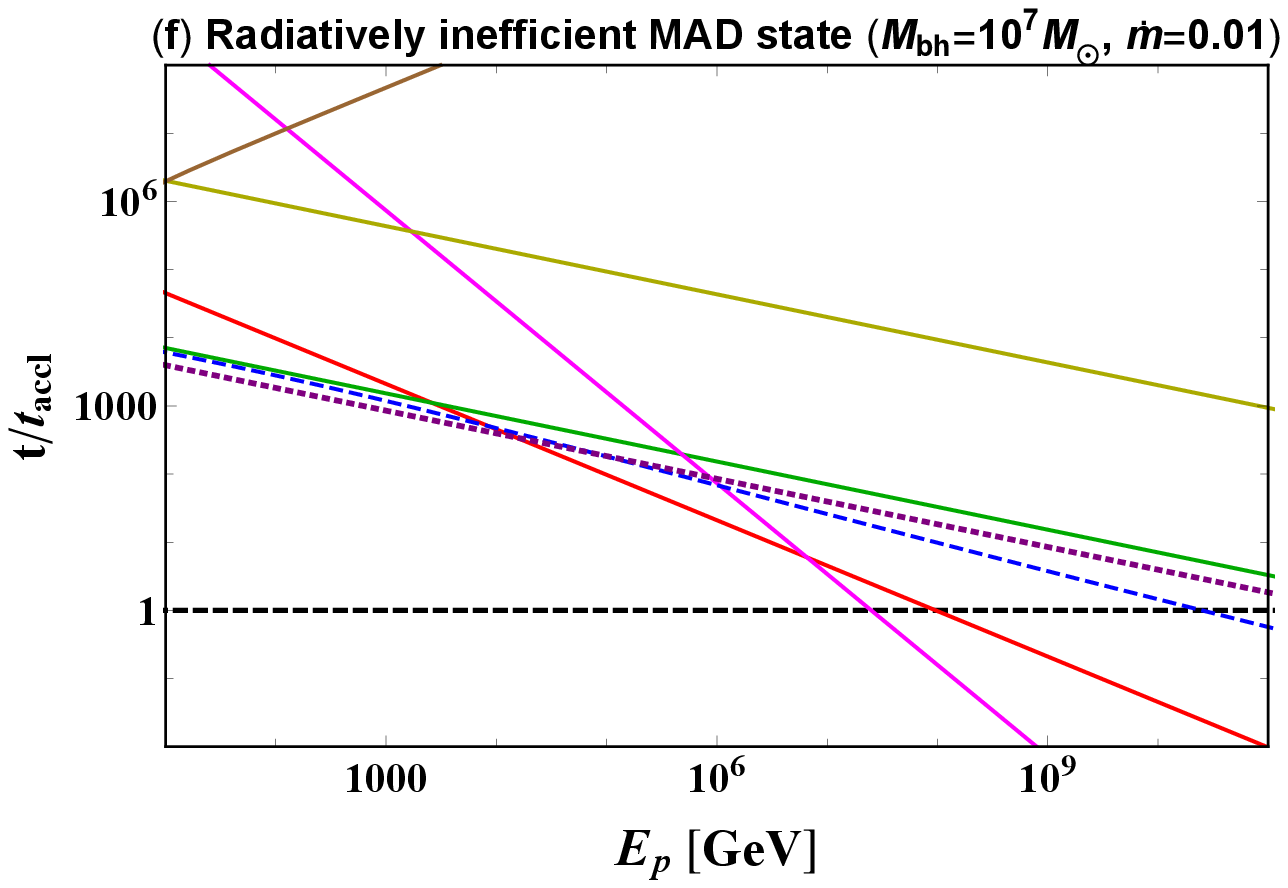}
\caption{
Dependence of the characteristic timescales normalized by the acceleration time 
on the proton energy for the six sites in the TDR. 
In each panel, the dashed black line denotes the time of stochastic acceleration of relativistic 
protons ($t_{\rm accl}$). The solid red, solid dotted, dotted magenta, 
dashed blue, dashed-dotted orange, solid purple, solid yellow, and solid brown 
lines are the diffusion time ($t_{\rm diff}$), infall 
time ($t_{\rm inf}$), synchrotron cooling time ($t_{\rm sync}$), proton-proton 
collision cooling ($t_{pp}$), photo-meson cooling time ($t_{\rm p\gamma}$), 
Compton drag time ($t_{\rm Cd}$), relaxation time ($t_{\rm rel}$), and Coulomb 
collision time ($t_{\rm Coul}$), respectively. Panels (a)-(f) show the cases of 
the first shock during debris circularization, super- and sub-Eddington accretion phases, 
RIAF phase, super-Eddington MAD state, and radiatively inefficient MAD state, respectively.
}
\label{fig:timescale}
\end{figure}

Panel (f) of Figure~\ref{fig:timescale} depicts the dependence 
of characteristic timescales normalized by $t_{\rm accl}$ on 
the proton energy in the radiatively inefficient MAD state. 
We note from the figure that the synchrotron process is 
the dominant cooling mechanism. From equation~(\ref{eq:gammasync}), 
we can estimate the Lorenz factor at $t_{\rm accel}=t_{\rm sync}$ as
\begin{eqnarray}
\gamma_{\rm sync}(r_{\rm p})
&\sim&
%1.3\times10^{8}\,
2.7\times10^7\,
\left(\frac{\epsilon}{0.01}\right)^{5/8}
%\left(\frac{\alpha}{0.1}\right)^{3/4}
\left(\frac{\zeta}{0.1}\right)^{3/4}
%\left(\frac{\mathcal{B}}{3}\right)^{-7/2}
\left(\frac{\beta}{1.0}\right)^{-5/16} 
\left(\frac{M_{\rm bh}}{10^7\,M_\odot}\right)^{-1/12} 
\nonumber \\
&\times&
\left(\frac{\dot{m}}{0.01}\right)^{-5/8} 
\left(\frac{m_*}{M_\odot}\right)^{-5/48} 
\left(\frac{r_*}{R_\odot}\right)^{5/16},
\label{eq:gammasync3}
\end{eqnarray}
where $s=5/3$ is adopted. The protons can be accelerated up to 
$E_{p,\rm {sync}}=\gamma_{\rm sync}m_pc^2\simeq25\,{\rm PeV}$.

%
%%%%%%%
% Figure 1
%%%%%%%
%
%

%
%%%%%%%
% Figure 2
%%%%%%%
%
\begin{figure}[!ht]
\centering
\includegraphics[width=8.1cm]{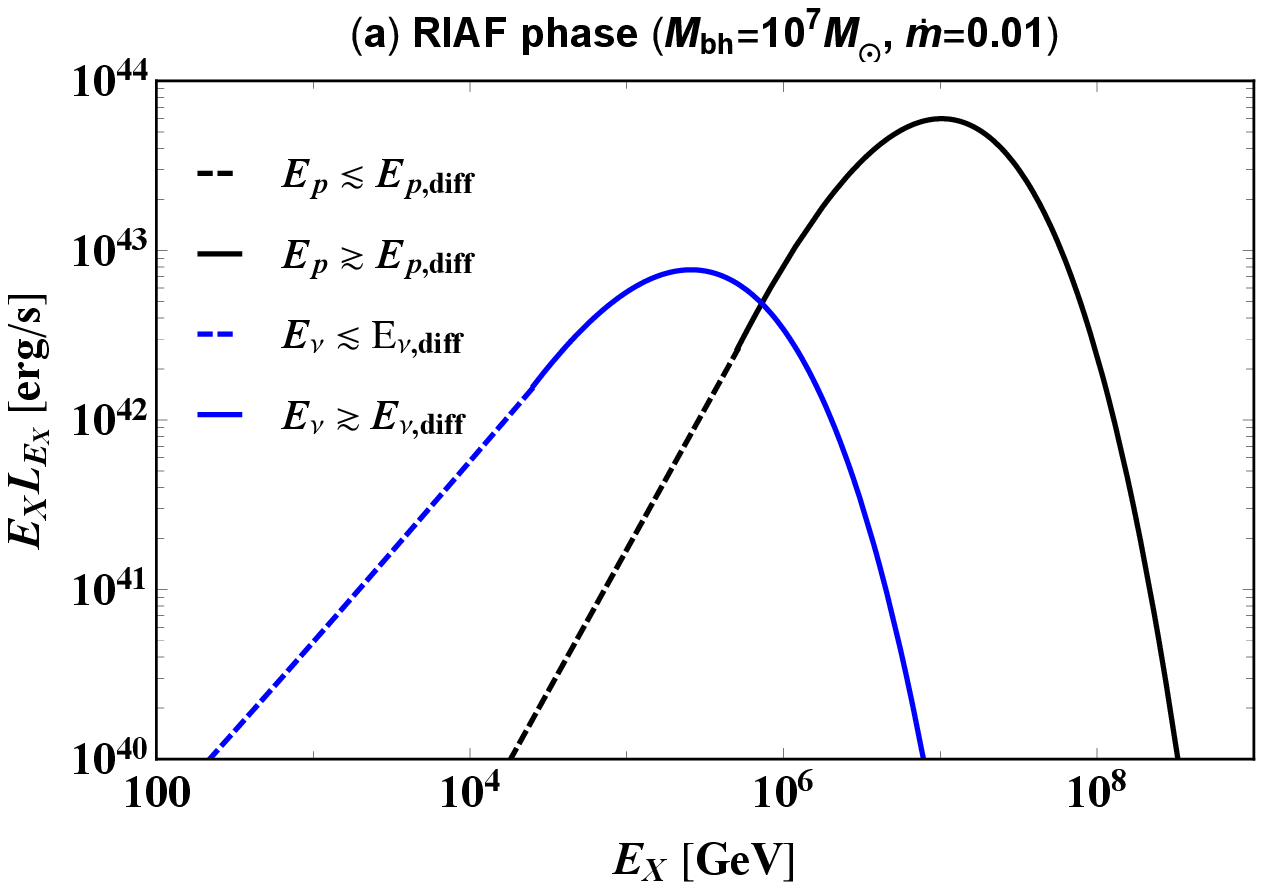}\\
\includegraphics[width=8.1cm]{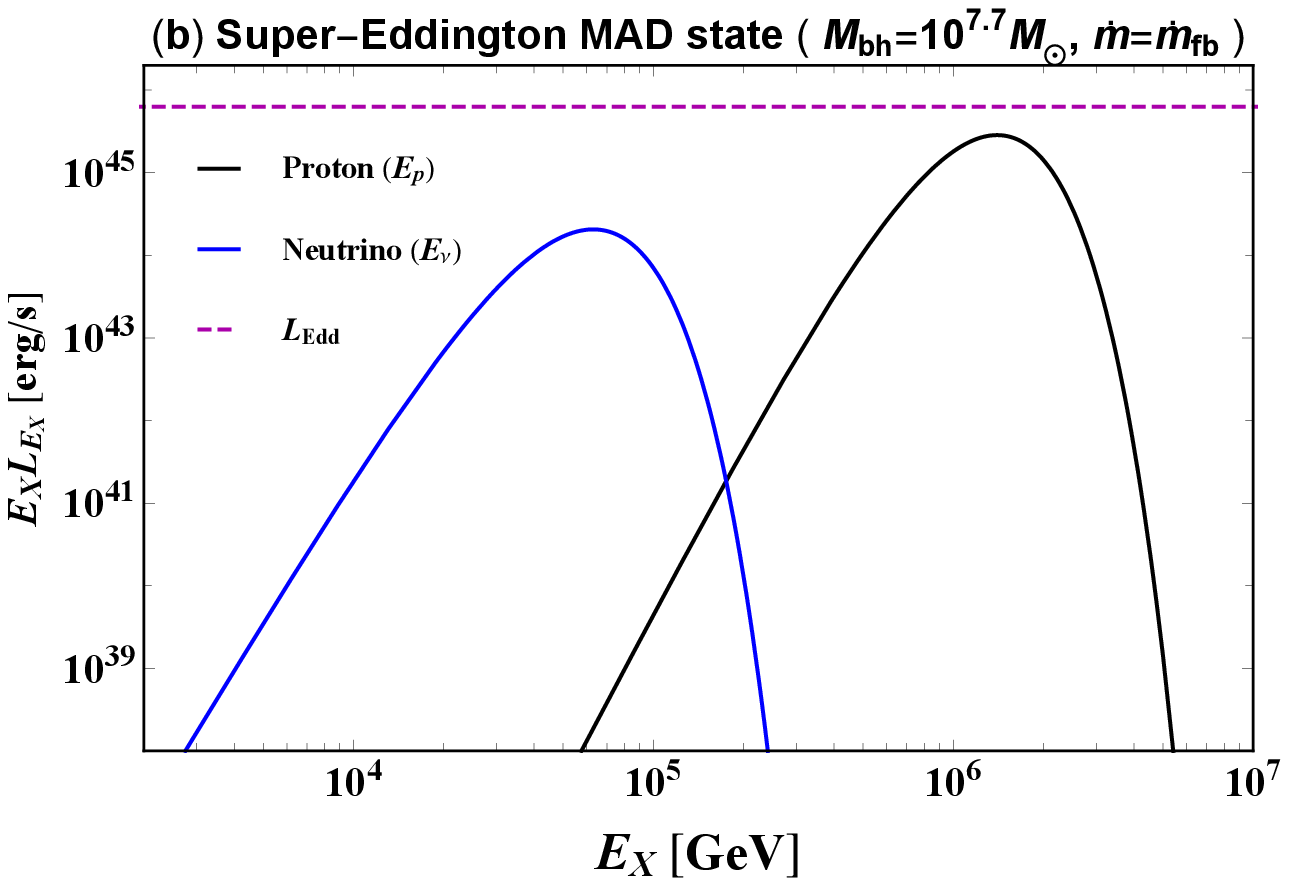}
\includegraphics[width=8.1cm]{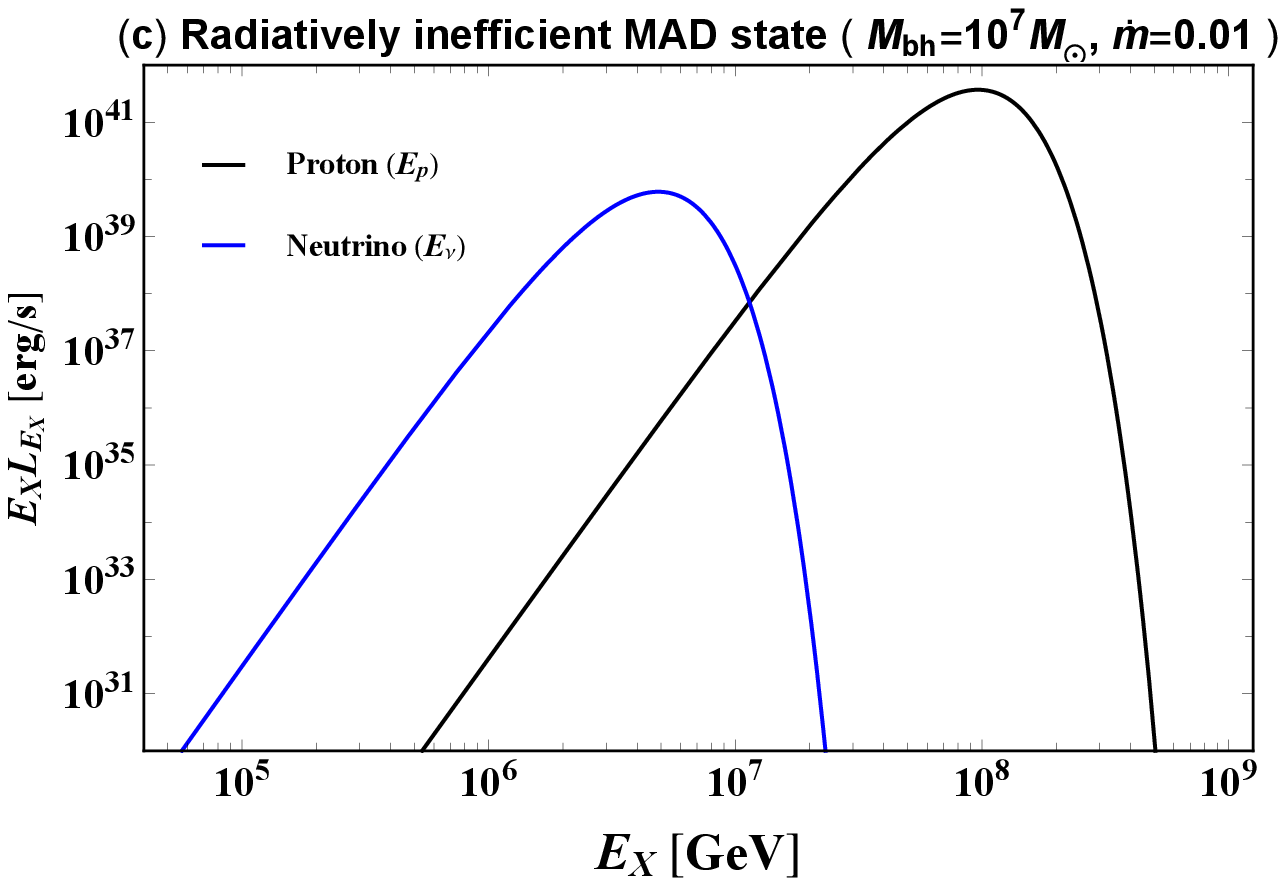}
\caption{
%Spectral energy distributions (SEDs) 
The differential luminosity spectra of the protons and neutrinos emitted from the RIAF phase 
(non-MAD state), super-Eddington MAD state, and radiatively inefficient MAD state. 
In all the three panels, the black and blue lines show the differential luminosity spectra 
of the proton and neutrino, respectively. 
Here $\dot{m}$, $E_{\rm p}$, and $E_{\nu}$ are the normalized mass fallback rate, proton energy, 
and neutrino energy, respectively, and $\eta_{\rm cr}=0.1$ is adopted as the injection efficiency. 
In panel (a), the dotted (solid) black and blue lines show the differential luminosity spectra of lower (higher) energy than 
$E_{p,{\rm diff}}$ (see equation \ref{eq:diffeq}) and $E_{\nu, {\rm diff}}=0.05E_{p,{\rm diff}}$, respectively.
Panel (b) depicts the differential luminosity spectra of super-Eddington MAD state, where the dashed line denotes the Eddington luminosity $L_{\rm Edd}\simeq6.3\times10^{45}\,{\rm erg/s}\,(M_{\rm bh}/10^{7.7}\,M_\odot)$. Panel (c) shows the case of the radiatively 
inefficient MAD state.
%The peak energy of the proton, $\gamma$-ray phorton, and neutrino are $E_{\rm p, pk}=(4913/216)E_{\rm{p, eq}}$, $E_{\rm p, pk}=(4913/2160)E_{\rm{p, eq}}$, and $E_{\rm \nu, pk}=(4913/4320)E_{\rm{p, eq}}$, respectively.
}
\label{fig:spectra}
\end{figure}

%
%
%%%%%%%%%%%%%%%%%%%%%%%%%%
\section{Neutrino Spectra and Luminosities}
\label{sec:3}
%%%%%%%%%%%%%%%%%%%%%%%%%%
%

The bolometric luminosity of the protons is defined by
\begin{eqnarray}
L_{p}\equiv\int\,dV\int\,dp\,\frac{4\pi{p^3}F(p)\,c}{t_{\rm diff}},
\label{eq:lep}
\end{eqnarray}
where $dV=4\pi{r}^2dr$ because the disk is assumed to be a spherically symmetric, 
$t_{\rm diff}$ is the shortest time when the acceleration is prevented in the RIAF phase, 
and $F(p)$ is the distribution function of the nonthermal protons. 
According to \citet{2006ApJ...647..539B}, $F(p)$ is given by
\begin{eqnarray}
F(p)=
\left\{ \begin{array}{ll}
A_{l}E_{p}^{-(1+s)}  & E_p  < E_{p,{\rm diff}} \\
A_{h}E_{p}^{-3/2}\,\exp\left[-\left(E_p/E_{p, {\rm diff}}\right)^{2-s}/(2-s)\right] & E_p \gtrsim E_{p,{\rm diff}}, \\
\end{array} \right.
\label{eq:alh}
\end{eqnarray}
where $A_l$ and $A_h$ are the normalization coefficients and 
$A_{\rm l}/A_{\rm h}=E_{p,{\rm diff}}^{(2s-1)/2}/e^{3}$ at $E_p=E_{p, \rm diff}$.
By substituting equations (\ref{eq:tdiff}) and (\ref{eq:alh}) into equation (\ref{eq:lep}) with the assumption that 
$L_{p}=\eta_{\rm cr}\dot{M}c^2$, $A_l$ is determined to be 
\begin{eqnarray}
A_{\rm l}=
\frac{(5-2s)(10-s)}{64\pi^2}\left(\frac{c}{r_{\rm d}}\right)^{3}
E_{p,{\rm diff}}^{s-3}\,t_{\rm diff}(r_{\rm d},\gamma_{\rm diff})\,
\eta_{\rm cr}\dot{M}c^2,
\nonumber
\end{eqnarray}
where $\eta_{\rm cr}$ is the injection efficiency for the protons, and we use $\eta_{\rm cr}=0.1$ 
as a fiducial value unless otherwise noted. Similarly, $A_{\rm h}$ is estimated to be 
\begin{eqnarray}
A_{\rm h}=\frac{(5-2s)(10-s)}{64\pi^2e^3}
\left(\frac{c}{r_{\rm d}}\right)^{3}E_{p,{\rm diff}}^{-5/2}\,t_{\rm diff}(r_{\rm d},\gamma_{\rm diff})\,
\eta_{\rm cr}\dot{M}c^2,
\nonumber
\end{eqnarray}
where 
$
t_{\rm diff}(r,\gamma_{\rm diff})\sim10^{4}\,{\rm s}\,
\left(
\mathcal{B}/3
\right)^{1/2}
\left(
r/r_{\rm d}
\right)^{3/2}
\left(
M_{\rm bh}/10^7\,M_\odot
\right)^{-1/2}
$
with $\sigma_{pp}\approx3.6\times10^{-26}\,{\rm cm^2}$.

%
% 2nd paragraph
%
The differential luminosity of the protons is estimated to be
\begin{eqnarray}
E_{p}L_{E_{p}}
&=&
\int{dV}\frac{4\pi{p}^3F(p)E_{\rm p}}{t_{\rm diff}}
=(5-2s)\,\eta_{\rm cr}\,\dot{M}c^2 
\nonumber \\
&\times&
\left\{ \begin{array}{ll}
\left(E_{p}/E_{p,{\rm diff}}\right)^{5-2s} & E_{p}  \lesssim E_{p,{\rm diff}} \\
\left(E_{p}/E_{p,{\rm diff}}\right)^{9/2-s}\,
\exp\left[(1-\left(E_{p}/E_{p,{\rm diff}}\right)^{2-s})/(2-s)\right] & E_{p} \gtrsim E_{p,{\rm diff}}. \\
\end{array} \right.
\label{eq:eplp}
\end{eqnarray}
The corresponding spectra of the neutrinos are defined by
\begin{eqnarray}
E_{\nu}L_{E_\nu}
&=&
\int{dV}\frac{4\pi{p}^3F(p)E_{p}}{t_{pp}}=
\eta_{\rm cr}\,\dot{M}c^2 
\times
\frac{t_{\rm diff}}{t_{pp}}
\times
\frac{(10-s)(5-2s)}{2}
\nonumber \\
&\times&
\left\{ \begin{array}{ll}
\left(E_{ \nu}/E_{\rm \nu,diff}\right)^{3-s} & E_{\nu}  \lesssim E_{\rm \nu, diff} \\
\left(E_{\nu}/E_{\rm \nu,diff}\right)^{5/2}\,
\exp\left[(1-\left(E_{\nu}/E_{\rm \nu,diff}\right)^{2-s})/(2-s)\right] & E_{\nu} \gtrsim E_{\rm \nu, diff}, \\
\end{array} \right.
\label{eq:enulnu}
\end{eqnarray}
where $E_{\nu}=0.05E_p$ (thus $E_{\nu,{\rm diff}}=0.05E_{p,{\rm diff}}$) 
and the ratio of diffusion to proton-proton collision timescales is
\begin{eqnarray}
\frac{t_{\rm diff}}{t_{pp}}
&=&\frac{9}{2}\zeta{n}_{\rm p}\sigma_{pp}
r_{\rm }^{3-s}
\left(\frac{r_{\rm L}}{m_{p}c^2}\right)^{s-2}
\left(\frac{E_p}{E_{p,{\rm diff}}}\right)^{s-2}
E_{p,{\rm diff}}^{s-2}.
\nonumber
\end{eqnarray}
By adopting $s=5/3$ and $\sigma_{pp}\approx3.6\times10^{-26}\,{\rm cm^2}$, 
we obtain 
$
t_{\rm diff}/t_{pp}
\sim
1.7\times10^{-2}
\left(\alpha/0.1\right)^{-1}
\left(\mathcal{B}/3\right)^{1/2}
\left(\dot{m}/0.01\right)
\left(E_{\nu}/E_{\rm \nu,diff}\right)^{-1/3}
$.
By differentiating equations (\ref{eq:eplp}) and (\ref{eq:enulnu}), 
the energy of protons and neutrinos at the peak of the spectra is given by 
$E_{p,{\rm pk}}=(4913/216)E_{p,{\rm diff}}$ and $E_{\rm \nu,pk}=(125/8)E_{\nu, {\rm diff}}$, 
respectively. 
The differential luminosity of the protons at the peak is then given by
\begin{eqnarray}
E_{p}L_{E_p}|_{\rm pk}&=&(5-2s)\eta_{\rm cr}\dot{M}c^2
\left(\frac{E_{p,{\rm pk}}}{E_{p,{\rm diff}}}\right)^{9/2-s}\,
\exp\left[\frac{1}{2-s}\left(1-\left(\frac{E_{\rm p,pk}}{E_{p, {\rm diff}}}
\right)^{2-s}\right)\right]
\nonumber \\
&=& g_{p}(s)\eta_{\rm cr}M_{\rm bh}\,\dot{m} \nonumber \\
&\propto&{t}^{-5/3},
\label{eq:lppk}
\end{eqnarray}
where $g_p(s)=(5-2s)(E_{p,{\rm pk}}/E_{p,{\rm diff}})^{9/2-s}
\exp[(1-(E_{p,{\rm pk}}/E_{p, {\rm diff}})^{2-s})/(2-s)]$. 
Similarly, the differential luminosity of the neutrinos at the peak is given by
\begin{eqnarray}
E_{\nu}L_{E_\nu}|_{\rm pk}&=&\frac{(10-s)(5-2s)\eta_{\rm cr}\dot{M}c^2}{2}
\left(\frac{E_{\rm \nu,pk}}{E_{\nu,{\rm diff}}}\right)^{5/2}\,
\exp\left[\frac{1}{2-s}\left(1-\left(\frac{E_{\rm \nu,pk}}{E_{\nu,{\rm diff}}}\right)^{2-s}\right)\right]
\nonumber \\
&\times&\frac{t_{\rm diff}}{t_{pp}}\Biggr|_{\gamma=\gamma_{\rm pk}} \nonumber \\
&\propto&
g_{\nu}(s)\eta_{\rm cr}\alpha^{-1}\mathcal{B}^{1/2}\,M_{\rm bh}\,\dot{m}^2
\nonumber \\
&\propto&{t}^{-10/3},
\label{eq:lnupk}
\end{eqnarray}
where $g_\nu(s)=(10-s)(5-2s)(E_{\nu,{\rm pk}}/E_{\nu,{\rm diff}})^{5/2}
\exp[(1-(E_{\nu,{\rm pk}}/E_{\nu, {\rm diff}})^{2-s})/(2-s)]/2$.
The neutrino energy emitted at the peak is estimated as 
$E_{\nu,{\rm pk}}=(25/32)E_{p,{\rm diff}}\simeq0.35\,{\rm PeV}\,(M_{\rm bh}/10^7M_\odot)^{5/3}$ 
with the other given appropriate parameters.

The gamma-ray photons are also produced by pionic decay with 
$E_{\gamma,{\rm pk}}=(25/16)E_{p,{\rm diff}}\sim0.7\,{\rm PeV}$. 
They naturally cause the pair production by the interaction with the 
photons having the larger energy than $E_{\rm rad}=(m_e{c}^2)^2/E_{\gamma,{\rm pk}}\sim3.7\times10^{-4}\,{\rm eV}$.
The optical depth is estimated to be $\tau_{\gamma}\sim\sigma_{\rm T}({t_{\rm inf}L_{\rm rad}/E_{\rm rad}})
/(\pi{r}_{p}^2)\sim2.5\times10^4$, where $L_{\rm rad}\sim10^{36}\,{\rm erg/s}$ is adopted as the radio 
luminosity of the ADAF model (e.g. see Figure~\ref{fig:timescale} of \citealt{2015ApJ...806..159K}). Since the 
gamma-ray photons cause the pair production very efficiently because of $\tau_{\gamma}\gg1$, 
the gamma-ray emission is unlikely to be observed during the RIAF phase.

%
%%%%%%%%%%%
%      MAD states   
%%%%%%%%%%%
%

Next, for the MAD state, the synchrotron radiation can be dominant among the other cooling processes 
because of the stronger magnetic field than the non-MAD case. In this case, the bolometric 
luminosity of the accelerated protons is given by  
\begin{eqnarray}
L_{p}=\int\,dV\int\,dp\,\frac{4\pi{p^3}F_{\rm sync}(p)\,c}{t_{\rm sync}},
\label{eq:les}
\end{eqnarray}
where the distribution function of protons, $F_{\rm sync}(p)$, for the synchrotron cooling case 
is given by \cite{2008ApJ...681.1725S} as
\begin{eqnarray}
F_{\rm sync}(p)=
A_{\rm s}\left(\frac{E_{\rm p}}{c}\right)^{2}\,\exp\left[-\frac{1}{3-s}\left(\frac{E_{\rm p}}{E_{\rm p,sync}}\right)^{3-s}\right]
\label{eq:sync}
\end{eqnarray}
with the normalization coefficient $A_{\rm s}$ for the entire energy range. 
By substituting equations (\ref{eq:tsync}) and (\ref{eq:sync}) into equation (\ref{eq:les}) 
with the assumption that $L_{p}=\eta_{\rm cr}\dot{M}c^2$, $A_{\rm s}$ is determined as
\begin{eqnarray}
A_{\rm s}=\frac{1}{32\pi^2}\frac{1}{g_s(s)}
\left(\frac{c}{r_{\rm d}}\right)^{3}\frac{c^2}{E_{\rm p, sync}^{6}}\,t_{\rm sync}(r_{\rm d},\gamma_{\rm sync}(r_{\rm d}))\,
\eta_{\rm cr}\dot{M}c^2,
\end{eqnarray}
where $g_{\rm s}(s)\equiv\int_{0}^{\infty}x^{6}\exp[-x^{1/(3-s)}/(3-s)]\,dx$ and $g_{\rm s}(5/3)\simeq120$.

The differential luminosity of the protons is estimated to be
\begin{eqnarray}
E_{p}L_{E_{p}}
=
\int{dV}\frac{4\pi{p}^3F_{\rm sync}(p)E_{p}}{t_{\rm sync}}
=
\frac{\eta_{\rm cr}\,\dot{M}c^2 
}{g_{\rm s}(s)}\left(\frac{E_p}{E_{p,{\rm sync}}}\right)^{7}\,
\exp\left[\frac{-1}{3-s}\left(\frac{E_p}{E_{p,{\rm sync}}}\right)^{3-s}
\right]
\label{eq:eplpmad}
\end{eqnarray}
for the entire energy range. 
The corresponding spectra of the neutrinos are defined by
\begin{eqnarray}
E_{\nu}L_{E_\nu}
\simeq
\int{dV}\frac{4\pi{p}^3F_{\rm sync}(p)E_{p}}{t_{pp}}
&=&
\frac{\eta_{\rm cr}\dot{M}c^2 }{3g_{\rm s}(s)}
\left(\frac{E_p}{E_{p,{\rm sync}}}\right)^{6}\,
\exp\left[\frac{-1}{3-s}\left(\frac{E_p}{E_{p,{\rm sync}}}\right)^{3-s}
\right] \nonumber \\
&\times&
\frac{t_{\rm sync}(r_{\rm d},\gamma_{\rm sync}(r_{\rm d}))}{t_{pp}(r_{\rm d})},
\label{eq:enulnumad}
\end{eqnarray}
where the ratio of synchrotron cooling to proton-proton collision timescales is given by
\begin{eqnarray}
\frac{t_{\rm sync}(r_{\rm d},\gamma_{\rm sync})}{t_{pp}(r_{\rm d})}
&=&
\frac{3}{4}K_{pp}\epsilon
\left(\frac{m_{p}}{m_{e}}\right)^{2}
\left(\frac{v_{\rm K}(r_{\rm d})}{c}\right)^{-2}
\frac{\sigma_{pp}}{\sigma_{\rm T}}
\left(\frac{H}{r}\right)^{-1}
\frac{1}{\gamma_{\rm sync}(r_{\rm d})}.
%\frac{9}{4}K_{pp}\mathcal{B}
%\left(\frac{m_{p}}{m_{e}}\right)^{2}
%\left(\frac{v_{\rm K}(r_{\rm d})}{c}\right)^{-2}
%\frac{\sigma_{pp}}{\sigma_{\rm T}}
%\frac{1}{\gamma_{\rm sync}(r_{\rm d})}.
%\nonumber 
%&\sim&
%1.7\times10^{-2}
%\left(\frac{\alpha}{0.1}\right)^{-1}
%\left(\frac{\mathcal{B}}{3}\right)^{1/2}\left(\frac{\dot{m}}{0.01}\right)\left(\frac{E_{\nu}}{E_{\rm \nu,sync}}\right)^{s-2}.
\end{eqnarray}
By differentiating equations (\ref{eq:eplpmad}) and (\ref{eq:enulnumad}), 
the energy of the protons and neutrinos at the peak of the spectra is given by 
$E_{p,{\rm pk}}=6^{3/4}E_{p,{\rm sync}}$ and $E_{\nu,{\rm pk}}=(6^{3/4}/20)E_{p,{\rm sync}}$, 
respectively. The differential luminosity of the protons at the peak is given by 
\begin{eqnarray}
E_{p}L_{E_p}|_{\rm pk}&=&\frac{\eta_{\rm cr}}{g_{\rm s}(s)}
\left(\frac{E_{p,{\rm pk}}}{E_{p,{\rm sync}}}\right)^{6}
\exp\left[-\frac{1}{3-s}\left(\frac{E_{p,{\rm pk}}}{E_{p,{\rm sync}}}\right)^{3-s}\right]
\,\dot{M}c^2 \nonumber \\
&\propto& M_{\rm bh}\dot{m} \nonumber \\
&\propto&{t}^{-5/3}.
\end{eqnarray}
Similarly, the neutrino luminosity at the peak is given by
\begin{eqnarray}
E_{\nu}L_{E_\nu}|_{\rm pk}&=&\frac{1}{3}\frac{\eta_{\rm cr}}{g_{\rm s}(s)}\left(\frac{E_{\rm \nu,pk}}{E_{p,{\rm sync}}}\right)^{6}
\exp\left[-\frac{1}{3-s}\left(\frac{E_{\nu,{\rm pk}}}{E_{\nu,{\rm sync}}}\right)^{3-s}\right]
\,\dot{M}c^2 
\times
\frac{t_{\rm sync}(r_{\rm d},\gamma_{\rm sync})}{t_{pp}(r_{\rm d})}
\nonumber \\
&\propto& 
\gamma_{\rm sync}^{-1}\,\mathcal{B}\,M_{\rm bh}\,\dot{m}
\propto
\zeta^{1/(s-3)}\epsilon^{(2-s)/(6-2s)}\alpha^{1/(s-3)}M_{\rm bh}^{(8-s)/(12-4s)}r_{\rm d}^{-s/(12-4s)}
\dot{m}^{(6-s)/(6-2s)}
\nonumber \\
&\propto&
t^{-(5/6)(6-s)/(3-s)}.
\end{eqnarray}
For $s=5/3$, $E_{\nu}L_{E_\nu}|_{\rm pk}$ is proportional to $t^{-65/24}$. 
This is steeper than the standard decay rate. 
For the super-Eddington MAD state, the neutrino energy at the peak is calculated 
to be $E_{\nu,{\rm pk}}=6^{3/4}E_{p,{\rm sync}}\simeq80\,{\rm TeV}\,(M_{\rm bh}/10^7M_\odot)^{41/48}$ 
with the other given appropriate parameters. 
As discussed in the last paragraph of Section~\ref{sec:semad}, the Compton drag is 
the most efficient mechanism to prevent the protons from accelerating in the super-Eddington 
MAD if $M_{\rm bh}\lesssim10^{7.7}\,M_\odot$.
In this case, $E_{\nu}L_{E_\nu}\simeq\int{dV}\,4\pi{p}^3F_{\rm Cd}(p)E_{p}/t_{pp}$, where 
$F_{\rm Cd}(p)\propto\,{p}^{2}\,\exp\left[-\left(E_p/E_{p,{\rm Cd}}\right)^{2-s}/(2-s)\right]$ 
is the distribution function of protons for the Compton drag \citep{2008ApJ...681.1725S}.
The neutrino luminosity is then estimated as $E_{\nu}L_{E_\nu}|_{\rm pk}\propto\eta_{\rm cr}\dot{M}c^2\propto{t^{-5/3}}$ 
at the peak of the neutrino energy $E_{\nu,{\rm pk}}=(54/5)\,E_{p,{\rm Cd}}\sim24\,{\rm TeV}\,(M_{\rm bh}/10^{7.4}\,M_\odot)^{89/12}$. The optical depth for Thomson scattering is estimated to be $\tau_{p}=2.9\times10^2\,(\epsilon/0.01)^{-1}(\beta/1.0)^{1/2}(M_{\rm bh}/10^{7.7}\,M_\odot)^{-7/6}(m_*/M_\odot)^{13/6}(r_*/R_\odot)^{-2}(\dot{m}/\dot{m}_{\rm fb})(r/r_{\rm p})^{-1/2}$. Because it is much larger than unity, the super-Eddington MAD is highly opaque 
to gamma-ray photons produced by pionic decay.

For the radiatively inefficient MAD state, the neutrino energy at the peak is calculated to be 
$E_{\nu,{\rm pk}}\simeq4.8\,{\rm PeV}\,(M_{\rm bh}/10^7M_\odot)^{-1/12}$ with the other given 
appropriate parameters. As well as what we discussed above for the case of the non-MAD state, 
the gamma-ray photons produced by pionic decay are unlikely to escape from the radiatively 
inefficient MAD because of the very efficient pair production.
%

%spectral energy distributions (SEDs)
%
Figure~\ref{fig:spectra} shows the differential luminosity spectra of the protons and neutrinos.
Panels (a)-(c) show the cases of the super-Eddington MAD state, the RIAF phase, 
and the radiatively inefficient MAD state, respectively. It is noted from the three panels that the neutrino 
emission has a nearly Eddington luminosity at the peak in the MAD state, while the neutrino emissions 
has $E_{\nu}L_{E_{\nu}}\sim7.0\times10^{42}\,{\rm erg\,s^{-1}}\,(\eta_{\rm cr}/0.1)$ at the peak in the 
RIAF case of the non-MAD state and $E_{\nu}L_{E_{\nu}}\sim3.2\times10^{39}\,{\rm erg\,s^{-1}}\,(\eta_{\rm cr}/0.1)$ 
at the peak in the RIAF case of the MAD state.
%

%
%%%%%%%%%%
\section{Discussion}
\label{sec:dis}
%%%%%%%%%%
%

In the MAD state, the magnetic field is so strong that the proton synchrotron radiation is efficient.
Since the energy spectrum of the relativistic proton is very hard (see equation~\ref{eq:eplpmad}), 
the typical photon energy is given by
\begin{eqnarray}
\langle h\nu_{\rm sync} \rangle &=&
\frac{hq_{\rm e}}{2\pi m_pc}B_{\rm MAD}(r_{\rm p}) \gamma_{\rm sync}(r_{\rm p})^2 
\sim0.34~{\rm MeV}
\left(\frac{B_{\rm MAD}(r_p)}{1.2\times10^6\,{\rm G}}\right)
\left(\frac{\gamma_{\rm sync}(r_p)}{4.5\times10^5}\right)^2,
\end{eqnarray}
where we use equations (\ref{eq:bmad2}) and (\ref{eq:gammasync}).
Such photons can potentially become the target for ${p}\gamma$ cooling.
Here we roughly estimate $t_{p\gamma}$ for such a case.
Since $L_{E_p}\propto E_p{}^5$ from equation (\ref{eq:eplpmad}), 
we approximate the energy density of proton synchrotron photons as 
$U_\gamma(E_\gamma)\sim U_{\gamma,{\rm pk}}(E_{\gamma}/\langle h\nu_{\rm sync} \rangle)^3$
for $E_{\gamma}<\langle h\nu_{\rm sync} \rangle$ and $U_\gamma(E_\gamma)\sim 0$
for $E_{\gamma}>\langle h\nu_{\rm sync} \rangle$.
For protons with energy $E_{p,{\rm sync}}=\gamma_{\rm sync}(r_p)m_pc^2$, almost all their energy is converted to
the synchrotron emission which is the dominant cooling process, so that we can write 
$U_{\gamma,{\rm pk}}\sim (r_p/c)L_p/\langle h\nu_{\rm sync} \rangle V$,
where $V\sim{r}_{\rm p}^3$ is the volume of the emission region. Then the photon number spectrum is written as
$N_\gamma(E_\gamma)\sim(U_{\gamma,{\rm pk}}/\langle h\nu_{\rm sync} \rangle)(E_{\gamma}/\langle h\nu_{\rm sync} \rangle)^2$
for $E_{\gamma}<\langle h\nu_{\rm sync} \rangle$.
Substituting this into equation~(\ref{eq:tpgamma}), we derive the ${p}\gamma$ cooling time for the 
protons as 
\begin{eqnarray}
t_{p\gamma}(\gamma)\sim10^{-5}~{\rm s} ~\gamma^2\left(1-\frac{\gamma_c}{\gamma}\right)^{-1} 
\left(\frac{\langle h\nu_{\rm sync} \rangle}{0.1~{\rm MeV}}\right)^3
 %~~~~~~~~~~~~ \times
%\left(\frac{L_p}{10^{45}~{\rm erg~s}^{-1}}\right)^{-1}
\left(\frac{L_p}{L_{\rm Edd}}\right)^{-1}
\left(\frac{r_p/c}{10^3~{\rm s}}\right)^{-1}
\left(\frac{V}{10^{39}~{\rm cm}^3}\right),
\end{eqnarray}
which is validated when $\gamma>\gamma_c=\bar{\epsilon}_{\rm pk}/2\langle h\nu_{\rm sync} \rangle\sim500$.
For protons with $\gamma<\gamma_c$, there are no target photons in their rest frame.
Namely, if $\gamma\gg\gamma_c$, one can see
$t_{p\gamma}\sim10^{5}(\gamma/10^5)^2$~s, which is much longer than the acceleration time $t_{\rm accl}$.
Therefore, the proton acceleration is not limited by the ${p}\gamma$ cooling.

The baryonic loading parameter is given by $\xi_{\rm bl}=L_{p}/L_{\rm \gamma}$, 
where $L_{p}=\eta_{\rm cr}\dot{M}_{\rm fb}c^2\sim2.4\times10^{45}\,{\rm erg\,s^{-1}}\,(\eta_{\rm cr}/0.1)
\left(M_{\rm bh}/10^{7.7}\,M_\odot\right)^{-1/2}(m_{*}/M_\odot)^2(r_*/R_\odot)^{-3/2}(t/t_{\rm mtb})^{-5/3}$ 
and $L_\gamma$ is the photon's luminosity.
Adopting the slim-disk model for the super-Eddington MAD, we estimate $L_{\gamma}=2\pi\int_{r_{\rm S}}^{r_{\rm t}}\,rQ_{\rm rad}\,dr\approx1.1\times10^{46}\,(M_{\rm bh}/10^{7.7}\,M_\odot)\log(r_{\rm t}/r_{\rm s})\,{\rm erg\,s^{-1}}$, where 
$Q_{\rm rad}\approx3.7\times10^{17}\,{\rm erg/cm^{2}}(M_{\rm bh}/10^{7.7}\,M_\odot)^{-1}(r/r_{\rm S})^{-2}$ 
is the radiative cooling rate \citep{2006ApJ...648..523W}. We then estimate $\xi_{\rm bl}\sim0.21\,(\eta_{\rm cr}/0.1)\left(M_{\rm bh}/10^{7.7}\,M_\odot\right)^{-3/2}(m_{*}/M_\odot)^2\\(r_*/R_\odot)^{-3/2}\log^{-1}(r_{\rm t}/r_{\rm s})$. 
Adopting $L_{\gamma}=\eta_{\rm MAD}\dot{M}c^2$ similarly, we estimate 
$\xi_{\rm bl}\sim0.67\,(\eta_{\rm cr}/0.1)(\eta_{\rm MAD}/0.15)^{-1}$. 
In the super-Eddington MAD state, the baryon loading is less than unity 
and therefore different from the jetted TDE case where 
$\xi_{\rm bl}$ ranges from $1$ to $100$ \citep{2015APh....62...66B,2017ApJ...838....3S,2018A&A...611A.101B}.
In the RIAFs, the baryonic loading parameter is estimated to be 
$\xi_{\rm bl}=L_{p}/L_{\rm RIAF}\sim1.0\,(\eta_{\rm cr}/0.1)(\dot{m}/0.01)(\alpha/0.1)^{-2}$, 
where we adopt $L_p=\eta_{\rm cr}\dot{M}c^2$ and $L_{\rm RIAF}\approx0.1(\dot{m}/\alpha^2)\dot{M}c^2$ \citep{1997ApJ...477..585M}. In this case, $\xi_{\rm bl}$ is of the order of unity or less in the RIAFs of both the non-MAD and MAD cases.

The neutrino energy generation rate inferred from the observed isotropic neutrino flux 
is estimated to be $\rho_{\nu}\sim10^{43-44}\,{\rm erg\,Mpc^{-3}\,yr^{-1}}$ for the $10-100\,{\rm TeV}$ range 
as seen in Figure~1 of \cite{2019PhRvD..99f3012M}. In our present model, the neutrino energy generation rate of the RIAF phase is estimated to be $\rho_\nu=L_{\nu,{\rm bol}}{t_{\rm RIAF}}R_{\rm V}\sim2.1\times10^{43}\,{\rm erg\,Mpc^{-3}\,yr^{-1}}\,(\eta_{\nu}/0.1)(\xi_{\rm bl}/1.0)(L_{\rm RIAF}/1.3\times10^{42}\,{\rm erg\,s^{-1}})(M_{\rm  bh}/10^7\,M_\odot)^{-3/5}(t_{\rm RIAF}/1.7\times10^{9}\,{\rm s})(\mathcal{R}_{\rm V}/10^{-7}\,{\rm Mpc^{-3}\,yr^{-1}})$, where we adopt $\dot{m}=0.01$ for $\dot{M}=\dot{m}\dot{M}_{\rm Edd}$, $L_{\nu,{\rm bol}}=\eta_{\nu}L_p=\eta_{\nu}\xi_{\rm bl}L_{\rm RIAF}$ is the bolometric luminosity of the neutrinos, $\eta_\nu\sim0.1$ is the proton to neutrino conversion efficiency, and $\mathcal{R}_{\rm V}$ is the volumetric TDE rate (cf. \citealt{2017MNRAS.469.1354D}). In the case of super-Eddington MAD, the energy generation rate is $\rho_{\nu}=L_{\nu,\rm{bol}}{t_{\rm Edd}}R_{\rm V}\sim2.3\times10^{43}\,{\rm erg\,Mpc^{-3}\,yr^{-1}}\,(\eta_{\rm cr}/0.1)(\xi_{\rm bl}/0.67)(L_{p}/6.3\times10^{45}\,{\rm erg\,s^{-1}})(M_{\rm  bh}/10^{7.7}\,M_\odot)^{-3/2}(t_{\rm mtb}/2.5\times10^{7}\,{\rm s})(\mathcal{R}_{\rm V}/10^{-9}\,{\rm Mpc^{-3}\,yr^{-1}})$, where we adopt $L_{\nu,{\rm bol}}=\eta_{\nu}L_p=\eta_{\nu}\xi_{\rm bl}L_{\rm Edd}$ and we assume that $\sim1\,\%$ of the observed TDE rate experiences the super-Eddington MAD state. The RIAFs of both the non-MAD and MAD states and the super-Eddington MAD state can potentially contribute to the diffuse neutrino flux. 
%
%However, the radiatively inefficient MAD case is unlikely to contribute to the diffuse neutrino flux because of the much smaller energy generation rate $\eta_\nu\sim1.0\times10^{-4}$ as seen in panel (c) of Figure~\ref{fig:spectra}.

%
It is interesting to refer to the redshift evolution of TDE rates.
According to \cite{2016MNRAS.461..371K}, the TDE rate rapidly 
decreases with redshift, mainly because the black hole mass density 
decreases with redshift. The TDEs that occurred at $z=0$ are the main 
source contributing to the diffuse neutrino flux. However, there is some 
ambiguity as to how the frequency of TDEs evolves with redshift because 
most of the identified TDEs occur at a lower redshift than unity. We will discuss 
in detail  how much the TDE remnants contribute to the diffuse neutrino flux 
in a forthcoming paper.
In the standard TDE theory, the RIAF phase would start at $t_{\rm RIAF}\sim10^9\,{\rm s}$ after 
the stellar disruption (see equation \ref{eq:triaf} for details). However, if a star approaches an 
SMBH on a marginally hyperbolic orbit, a small fraction of debris mass should have negative 
binding energy and therefore falls back to the SMBH at a much smaller rate than the Eddington 
rate \citep{2018ApJ...855..129H}. In this case, the RIAF phase starts at about two orders of 
magnitude earlier than the standard case, that is, $\sim10^{7}\,{\rm s}$. This can enhance the 
energy generation rate, even if the event rate of marginally hyperbolic TDEs would be subdominant.
The heavier nuclei exist for a main-sequence star with mass greater than 
$1\,M_\odot$ as well as for a WD case. Tidal disruption of such a massive 
star can show a significant enhancement of the nitrogen-to-carbon ratio due to 
the CNO cycle \citep{2016MNRAS.458..127K,2018ApJ...857..109G}.
These carbon/nitrogen anomalies should, therefore, be observational evidence for 
TDEs caused by the massive stars. The increase of the nucleon number density is 
responsible for the larger energy density of the plasma, resulting in higher energy of 
accelerated protons. This excess energy possibly helps one to diagnose whether a 
disrupted star originates from the massive stars on the main sequence.
When a star is tidally disrupted, about a half of the stellar debris is unbound and 
ejected from the system. It collides with an ambient matter, forming a shock 
to produce nonthermal radiations like supernova remnants \citep{2016ApJ...822...48G}. 
Such electromagnetic afterglows are potentially observed after the neutrino emission is 
detected because the peak time of the afterglows is $10^{3-4}\,{\rm yr}$ depending on 
parameters such as the ambient density.
Let us discuss how one can observationally identify the neutrino emissions 
from the TDEs based on our models. The neutrino horizon is defined as $D_{\nu}=\sqrt{L_\nu/(4\pi\mathcal{S})}$, 
because $L_\nu/4\pi{d}^2$, where $d$ is the distance between the earth and the source, 
should be greater than or equal to the sensitivity $\mathcal{S}$ for the neutrino detection.
The IceCube sensitivity of neutrinos below 100 TeV is roughly given by $5\times10^{-11}\,{\rm erg\,s^{-1}\,cm^2}$ 
for an observation of a few months \citep{2019arXiv190205792I}. 
We can evaluate the horizon of $0.1-100\,{\rm TeV}$ neutrinos as $D_\nu\approx1\,{\rm Gpc}\,(L_{\nu}/L_{\rm Edd})^{1/2}(\mathcal{S}/[5\times10^{-11}\,{\rm erg\,s^{-1}\,cm^{-2}}])^{-1/2}$ 
in the case of $M_{\rm bh}=10^{7.7}\,M_\odot$. We then estimate the detection rate of the MAD state as $\sim3.4\times10^{-1}\,\rm{yr^{-1}}(n_{\rm gal}/0.003\,{\rm Mpc}^{-3})(D_{\nu}/220\,{\rm Mpc})^3(\mathcal{R}/10^{-5}\,{\rm yr}^{-1})$, where we adopt $L_{\nu}={E}_{\nu,{\rm pk}}L_{E_{\nu,{\rm pk}}}\sim3.0\times10^{44}\,{\rm erg\,s^{-1}}\,(\eta_{\rm cr}/0.1)$ (see panel (b) of Figure~\ref{fig:spectra}) and $\mathcal{R}=10^{-5}\,{\rm yr^{-1}}$ as the TDE rate per galaxy. In the case of RIAF phase, we can similarly evaluate the horizon as $D_\nu\approx460\,{\rm Mpc}\,(L_{\nu}/L_{\rm Edd})^{1/2}(\mathcal{S}/[5\times10^{-11}\,{\rm erg\,s^{-1}\,cm^{-2}}])^{-1/2}$ in the case of $M_{\rm bh}=10^{7}\,M_\odot$. The detection rate is then estimated to be $\sim1.2\times10^{-3}\,\rm{yr^{-1}}(n_{\rm gal}/0.003\,{\rm Mpc}^{-3})(D_{\nu}/34\,{\rm Mpc})^3(\mathcal{R}/10^{-5}\,{\rm yr}^{-1})$, where we adopt $L_{\nu}={E}_{\nu,{\rm pk}}L_{E_{\nu,{\rm pk}}}\sim7.0\times10^{42}\,{\rm erg\,s^{-1}}(\eta_{\rm cr}/0.1)$ 
(see panel (a) of Figure~\ref{fig:spectra}).

A star approaches an SMBH on a Keplerian orbit, and it is tidally disrupted around the pericenter, where 
the gravitational wave (GW) can be also emitted with a burst-like variation \citep{2004ApJ...615..855K}. 
The GW frequency is then given by $f=2\sqrt{GM_{\rm bh}/r_{\rm p}^3}\sim1.3\times10^{-3}\,{\rm Hz}\,(\beta/1)^{3/2}(m_*/M_\odot)^{1/2}(r_*/R_\odot)^{-3/2}$. We evaluate the GW horizon $D_{\rm gw}$ of the TDE case using the quadruple formula as $D_{\rm gw}=(1/2)(m_*/M_{\rm bh})(r_{\rm S}/r_{\rm p})(r_{\rm S}/h)\sim9.4\,{\rm Mpc}\,(\beta/1)(h/10^{-21})^{-1}(M_{\rm  bh}/10^7M_\odot)^{2/3}(m_*/M_\odot)^{4/3}(r_*/R_\odot)^{-1}$, where $h$ is the GW amplitude. $LISA$, $DECIGO$, and $BBO$ can detect the amplitude $h\gtrsim10^{-21}$ for the frequency range $10^{-4}\,{\rm Hz}\lesssim{f}\lesssim0.1\,{\rm Hz}$, $h\gtrsim10^{-22}$ for the frequency range $f\gtrsim10^{-2}\,{\rm Hz}$, and $h\gtrsim10^{-22}$ for the frequency range $f\gtrsim10^{-3}\,{\rm Hz}$, respectively \citep{2015CQGra..32a5014M}. The detection rate is conservatively estimated to be $\sim2.5\times10^{-5}\,\rm{yr^{-1}}(n_{\rm gal}/0.003\,{\rm Mpc}^{-3})(D_{\rm gw}/9.4\,{\rm Mpc})^3(\mathcal{R}/10^{-5}\,{\rm yr}^{-1})$ for the $LISA$ range. If we adopt $\beta=4$ for a solar-type star, then $f\gtrsim10^{-3}\,{\rm Hz}$ and thus $D_{\rm gw}\sim380\,{\rm Mpc}(M_{\rm  bh}/10^7M_\odot)^{2/3}$ with $h=10^{-22}$. The detection rate is conservatively $0.1\,{\rm yr^{-1}}(n_{\rm gal}/0.003\,{\rm Mpc}^{-3})(\mathcal{R}_{\beta}/10^{-6}\,{\rm yr}^{-1})$ for the $DECIGO$ and $BBO$ range, where we simply define $\mathcal{R}_\beta=\mathcal{R}/\beta^2$ because the cross section between the SMBH and the star is proportional to $1/\beta^2$. \cite{nra13} derived the GW amplitude as having a steeper $\beta$ dependence, 
and suggested the tidal disruption of a WD by an IMBH is a more promising GW source candidate for the advanced LIGO bands, although the event rate of WD-IMBH disruptions still includes a large ambiguity. From the viewpoints of the multi-messenger observations, there should be a time lag between the GW and neutrino detections because of the different emission site and mechanism. The GWs arise from the strong compression of the star at the tidal disruption, while the neutrinos can emit in the super-Eddington MAD state, radiatively inefficient MAD state, and the RIAF phase after the debris circularization phase. Therefore, the GW signals and associated X-ray flares \citep{2004ApJ...615..855K,2009ApJ...705..844G} could be precursors of neutrino and corresponding photon emissions in the two states and the RIAF phase.

%
%%%%%%%%%%%%%%%%%
\section{Conclusions}
\label{sec:con}
%%%%%%%%%%%%%%%%%
%
We have investigated the high-energy emissions from the stellar 
debris moving around SMBHs in TDEs. There are four main 
evolutionally phases after the tidal disruption of a star: the debris 
circularization phase, the super-Eddington accretion phases of both 
the MAD and non-MAD states, the sub-Eddington accretion phase, 
and the RIAF phases of both the MAD and non-MAD states, 
respectively (see equation~\ref{eq:tdeclass}). We have found that 
there are three promising sites, where the high-energy particles 
can be produced from the TDR: the super-Eddington accretion 
phase of the MAD state and the RIAF phases of both the non-MAD 
and MAD states. Our main conclusions are summarized as follows:

\begin{enumerate}
\item 
High-energy particles at the first (strongest) shock during the debris 
circularization phase are unlikely to be produced, 
because the mass fallback rate exceeding the Eddington rate makes 
both the Compton drag and the proton-proton collision cooling very efficient.
Since the shocked region is far from the black hole, the magnetic field is 
relatively weak. The weak magnetic field is another reason why the 
stochastic acceleration of the protons becomes inefficient.

\item 
High-energy particles during the super-Eddington accretion phase of the non-MAD state 
are unlikely to be produced because of very efficient proton-proton collision cooling and 
Compton drag, even though the stochastic acceleration of the protons is more efficient 
because of the much stronger magnetic field than that of the first shock case.

\item Neutrinos and gamma-ray photons are produced by pions not only in the 
super-Eddington MAD scenario but also in the RIAF scenarios of both the non-MAD 
and MAD cases. However, the gamma-ray photons cannot escape from the source 
because of the large optical depth in the super-Eddington MAD state and of the very 
efficient pair production in the RIAF cases.

\item In the RIAF phase, the  protons can be accelerated up to 
$\sim0.45\,{\rm PeV}\,(M_{\rm bh}/10^7\,M_\odot)^{5/3}$ with the other 
given appropriate parameters. The neutrino energy emitted at the peak 
of the spectrum is then estimated as $\sim0.35\,{\rm PeV}\,(M_{\rm bh}/10^7M_\odot)^{5/3}$. 
The neutrino emission is proportional to $t^{-10/3}$, which is steeper 
than the standard TDE decay rate. This indicates that one can potentially identify whether 
a RIAF that existed around an SMBH has a TDE origin or not.

\item In the super-Eddington MAD state, the stronger magnetic field than 
in the non-MAD case makes it possible to accelerate the protons up to 
ultrarelativistic energies. 
The dominant process to prevent the protons from accelerating 
strongly depends on the black hole mass. If the black hole 
mass is larger than $\sim10^{7.7}\,M_\odot$, the synchrotron cooling 
is the most dominant process. In this case, the protons are accelerated up to 
an energy of $\sim0.35\,{\rm PeV}\,(M_{\rm bh}/10^{7.7}M_\odot)^{41/48}$ 
with the other given appropriate parameters. The neutrino energy at the peak of the 
spectrum is then estimated as $\sim67\,{\rm TeV}\,(M_{\rm bh}/10^7M_\odot)^{41/48}$. 
Otherwise, the Compton drag is the more efficient process. In this case, the resultant 
energy of the protons increases at most up to 
$\sim2.2\,{\rm TeV}\,(M_{\rm bh}/10^{7.4}M_\odot)^{89/12}$.

\item In the super-Eddington MAD state, if $M_{\rm bh}\gtrsim10^{7.7}\,M_\odot$, the neutrino light 
curve is proportional to $t^{-65/24}$, which is steeper than the standard TDE $t^{-5/3}$ decay rate.
On the other hand, it follows the $t^{-5/3}$ decay rate with energy more than $1\,{\rm TeV}$ 
if $10^{7.4}\,M_\odot\lesssim{M}_{\rm bh}<10^{7.7}\,M_\odot$. 
In both cases, the neutrino luminosity is of the order of the Eddington limit or the luminosity 
exceeding it. Such a high neutrino luminosity and characteristic light curve give us the ability 
to judge whether the TDE disk is in the MAD state or not.

\item In the radiatively inefficient MAD state, the stronger magnetic field than in the non-MAD case 
makes it possible to accelerate the protons up to $25\,{\rm PeV}\,(M_{\rm bh}/10^7M_\odot)^{-1/12}$. 
The neutrino energy estimated at the peak of the spectrum is then 
$4.8\,{\rm PeV}\,(M_{\rm bh}/10^7M_\odot)^{-1/12}$.
The resultant neutrino luminosity is, however, too weak to be detected with the current sensitivity of IceCube.
\end{enumerate}
%
%\appendix
%
%%%%%%%%%%%%%%
\section*{Acknowledgments}
%%%%%%%%%%%%%%
%
The authors thank the anonymous referee for fruitful comments and suggestions. 
The authors also thank Abraham Loeb and Kohta Murase for their helpful comments and suggestions.
This research has been supported by Basic Science Research Program through the National 
Research Foundation of Korea (NRF) funded by the Ministry of Education (NRF-2017R1D1A1B03028580 
to K.H.) and also by JSPS KAKENHI (Grant No.18H01232 to R.Y.).
%

%
%%%%%%%%%%%%%


\begin{thebibliography}{}
%%%%%%%%%%%%%
%
% Combined fit of spectrum and composition data as measured by the Pierre Auger Observatory
\bibitem[Aab et al.(2017)]{2017JCAP...04..038A} 
Aab, A., Abreu, P., Aglietta, M., et al.\ 2017, Journal of Cosmology and Astro-Particle Physics, 2017, 38
% 	
% First Observation of PeV-Energy Neutrinos with IceCube
\bibitem[Aartsen et al.(2013)]{2013PhRvL.111b1103A} 
Aartsen, M.~G., Abbasi, R., Abdou, Y., et al.\ 2013, Physical Review Letters, 111, 021103 
%
%
\bibitem[Abramowicz et al.(1988)]{1988ApJ...332..646A} 
Abramowicz, M.~A., Czerny, B., Lasota, J.~P., \& Szuszkiewicz, E.\ 1988, \apj, 332, 646 
%
%
%%%\bibitem[Acharya et al.(2013)]{2013APh....43....3A} 
%%%Acharya, B.~S., Actis, M., Aghajani, T., et al.\ 2013, Astroparticle Physics, 43, 3 
%
%\bibitem[Aarseth et al.(1974)]{aa74}
%Aarseth, S. J., Henon, M., Wielen, R., 1974, A\&A, 37, 183
%
% Dynamics of binary-disk interaction. 1: Resonances and disk gap sizes
%\bibitem[Artymowicz \& Lubow(1994)]{al94}
%Artymowicz P., Lubow S. H., 1994, ApJ, 421, 651
%
%%%\bibitem[\protect\citeauthoryear{Aharonian}{2000}]{2000NewA....5..377A} 
%%%Aharonian F.~A., 2000, NewA, 5, 377
%
% Ultrahigh-energy cosmic rays from tidally-ignited white dwarfs
\bibitem[Alves Batista \& Silk(2017)]{2017PhRvD..96j3003A} 
Alves Batista, R., \& Silk, J.\ 2017, \prd, 96, 103003 
%
% Discovery of an Outflow from Radio Observations of the Tidal Disruption Event ASASSN-14li
\bibitem[Alexander et al.(2016)]{2016ApJ...819L..25A} 
Alexander, K.~D., Berger, E., Guillochon, J., Zauderer, B.~A., \& Williams, P.~K.~G.\ 2016, \apjl, 819, L25
%
% A Continuum of H- to He-rich Tidal Disruption Candidates With a Preference for E+A Galaxies
\bibitem[Arcavi et al.(2014)]{ar14}
Arcavi, I., Gal-Yam, A., Sullivan, M., et al. 2014, ApJ, 793, 38
%
% New Physical Insights about Tidal Disruption Events from a Comprehensive Observational Inventory at X-Ray Wavelengths
\bibitem[Auchettl et al.(2017)]{2017ApJ...838..149A} 
Auchettl, K., Guillochon, J., \& Ramirez-Ruiz, E.\ 2017, \apj, 838, 149 
%
%
%\bibitem[Bahcall(1976)]{bw76}
%Bahcall, J. N., Wolf, R. A. 1976, ApJ, 209, 214
%
%\bibitem[Bardeen \& Petterson(1975)]{jj75} 
%Bardeen.~J.~M., Petterson,~J.~A., 1975, ApJ. 195, L65
%
% Rotating Black Holes: Locally Non-rotating Frames, Energy Extraction, and Scalar Synchrotron Radiation
%\bibitem[Bardeen et al.(1972)]{jws72} 
%Bardeen,~J.~M., Press,~W.~H., Teukolsky,~S.~A., 1972, ApJ. 178, 347
%
%\bibitem[Bate et al.(1995)]{bate95}
%Bate,~M. R., Bonnel,~I. A., Price,~N.M., 1995, MNRAS, 277, 362
%
%\bibitem[Bate(1995)]{mb95}
%Bate,~M. R., 1995, Ph.D. Thesis, ch.2, 30-31
%
%\bibitem[Baumgardt & Makino(2003)]{bm03} 
%Baumgardt, H., & Makino, J. 2003, MNRAS, 340, 227
%\bibitem[Baumgardt et al.(2004)]{bme04} 
%Baumgardt, H., Makino, J., \& Ebisuzaki, T. 2004, ApJ, 613, 1133 
%Baumgardt, H., et al. 2003a, ApJ, 582, L21
%???. 2003b, ApJ, 589, L25
%
%\bibitem[Benz (1990)]{benz90a}
%Benz W., 1990, in the Numerical Modeling of Nonlinear Steller Pulsations: Problems and Prospects, ed. Buchler,~R.J. (Dordrecht: Kluwer Academic Publishers), 269 
%
%\bibitem[Benz et al. (1990)]{benz90b}
%Benz,~W., Bowers,~R.L., Cameron,~A.G.W., Press,~W.H., 1990, ApJ, 348, 647
%
%\bibitem[Berczik et al.(2011)]{ber11}
% Berczik, P., Nitadori, K., Zhong, S., Spurzem, R., Hamada, T., Wang, X., Berentzen, I., Veles, A., and Ge, W. 2011, in International conference on High Performance Computing, Kyiv, Ukraine, October 8-10, 2011., p. 8-18
%
%\bibitem[Berczik et al.(2013a)]{ber13a} 
% Berczik, P., Spurzem, R., Wang, L., Zhong, S., Huang, S.\ 2013, Third International Conference ``High Performance Computing'', HPC-UA 2013, p.~52-59, 52
%
%\bibitem[Berczik et al.(2013b)]{ber13b}
%Berczik, P., Spurzem, R., Zhong, S., Wang, L., Nitadori, K., Hamada, T., Veles, A. 2013, n Lecture Notes in Computer Science, Vol. 7905, Procs. of 28th Intl. Supercomputing Conf. ISC 2013, Leipzig, Germany, June 16-20, 2013., ed. J. M. Kunkel, T. Ludwig, and H. E. Meuer (Springer Vlg.), 13-25
%
%\bibitem[Berti \& Volonteri(2008)]{bv08} 
%Berti, E., Volonteri, M., 2008, ApJ, 684, 822 
%
% Are gamma-ray bursts the sources of ultra-high energy cosmic rays?
\bibitem[Baerwald et al.(2015)]{2015APh....62...66B} 
Baerwald, P., Bustamante, M., \& Winter, W.\ 2015, Astroparticle Physics, 62, 66
%
\bibitem[Becker et al.(2006)]{2006ApJ...647..539B} 
Becker, P.~A., Le, T., \& Dermer, C.~D.\ 2006, \apj, 647, 539 
%
%
% Tidally disrupted stars as a possible origin of both cosmic rays and neutrinos at the highest energies
\bibitem[Biehl et al.(2018)]{2018NatSR...810828B}
Biehl, D., Boncioli, D., Lunardini, C., \& Winter, W.\ 2018, Scientific Reports, 8, 10828 
%
%
%Cosmic ray and neutrino emission from gamma-ray bursts with a nuclear cascade
\bibitem[Biehl et al.(2018)]{2018A&A...611A.101B} 
Biehl, D., Boncioli, D., Fedynitch, A., et al.\ 2018, \aap, 611, A101
%
%
\bibitem[\protect\citeauthoryear{Bisnovatyi-Kogan \& Ruzmaikin}{1974}]{1974ApSS..28...45B} 
Bisnovatyi-Kogan G.~S., Ruzmaikin A.~A., 1974, Ap\&SS, 28, 45
%
% Gravitational Radiation from Post-Newtonian Sources and Inspiralling Compact Binaries
%\bibitem[Blanchet(2006)]{lb06} 
%Blanchet,~L., 2006, Living. Rev. Relativity, 9, 4
%
% Post-Newtonian hydrodynamics and post-Newtonian GW generation from numerical relativity
%\bibitem[Blanchet et al.(1990)]{ltg90}
%Blanchet,~L., Damour,~T., Schafer,~G., 1990, MNRAS, 242, 289
%
%\bibitem[Blandford \& Znajek(1977)]{rr77}
%Blandford,~R.~D., Znajek,~R.~L.,,1977, MNRAS, 179, 433
%
%A Possible Relativistic Jetted Outburst from a Massive Black Hole Fed by a Tidally Disrupted Star
%\bibitem[Bloom et al.(2011)]{jsb+11}
%Bloom~J. S. et al., 2011, Science, 333, 203
%
%\bibitem[Bogdanovi{\'c} et al.(2014)]{BCA14} 
%Bogdanovi{\'c}, T., Cheng, R.~M., Amaro-Seoane, P., 2014, ApJ, 788, 99
%
\bibitem[Bonnerot et al.(2016)]{2016MNRAS.455.2253B}
Bonnerot, C. et al, 2016, MNRAS, 455, 2253 
%
% Long-term stream evolution in tidal disruption events
%
\bibitem[Bonnerot et al.(2017)]{2017MNRAS.464.2816B} 
Bonnerot, C., Rossi, E.~M., \& Lodato, G.\ 2017, \mnras, 464, 2816 
%
%
%\bibitem[Brockamp et al.(2011)]{bbk11}
%Brockamp, M., Baumgardt, H., Kroupa, P. 2011, MNRAS, 418, 1308
%
\bibitem[Brown et al.(2015)]{2015MNRAS.452.4297B}
Brown, G.~C., Levan, A.~J., Stanway, E.~R., et al.\ 2015, \mnras, 452, 4297 
%
%Relativistic jet activity from the tidal disruption of a star by a massive black hole
\bibitem[Burrows et al.(2011)]{2011Natur.476..421B} 
Burrows, D.~N., Kennea, J.~A., Ghisellini, G., et al.\ 2011, \nat, 476, 421 
%
% Swift J2058.4+0516: Discovery of a Possible Second Relativistic Tidal Disruption Flare?
\bibitem[Cenko et al.(2012)]{2012ApJ...753...77C} 
Cenko, S.~B., Krimm, H.~A., Horesh, A., et al.\ 2012, \apj, 753, 77 
%
% THE ULTRAVIOLET-BRIGHT, SLOWLY DECLINING TRANSIENT PS1-11af AS A PARTIAL TIDAL DISRUPTION EVENT
\bibitem[Chornock et al.(2014)]{2014ApJ...780...44C} 
Chornock, R., Berger, E., Gezari, S., et al.\ 2014, \apj, 780, 44 
%
%\bibitem[Coughlin, et al.(2016)]{eric16}
%Coughlin, E.R. et al. 2016, MNRAS, 455, 3612
%
% VARIABILITY IN TIDAL DISRUPTION EVENTS: GRAVITATIONALLY UNSTABLE STREAMS
%\bibitem[Coughlin \& Nixon(2015)]{2015ApJ...808L..11C} 
%Coughlin, E.~R., \& Nixon, C.\ 2015, \apjl, 808, L11
%
%\bibitem[Cannizzo et al.(1990)]{clg90} 
%Cannizzo, J.~K., Lee, H.~M., Goodman, J., 1990, ApJ, 351, 38 
%
%
\bibitem[Dai et al.(2015)]{2015ApJ...812L..39D} 
Dai, L., McKinney, J.~C., \& Miller, M.~C.\ 2015, \apj, 812, L39
%
%
% Can tidal disruption events produce the IceCube neutrinos?
\bibitem[Dai \& Fang(2017)]{2017MNRAS.469.1354D}
Dai, L., \& Fang, K.\ 2017, \mnras, 469, 1354 
%
% A Unified Model for Tidal Disruption Events
\bibitem[Dai et al.(2018)]{2018ApJ...859L..20D} 
Dai, L., McKinney, J.~C., Roth, N., et al.\ 2018, \apj, 859, L20.
%
%
%Third post-Newtonian dynamics of compact binaries: Noetherian conserved quantities and equivalence between the harmonic-coordinate and ADM-Hamiltonian formalisms
%\bibitem[de~Andrade et al.(2001)]{vlg01} 
%de Andrade,~V.~C., Blanchet,~L., Faye,~G., 2001, Class. Quantum Grav, 18, 753
%
% Stochastic Particle Acceleration near Accreting Black Holes
\bibitem[Dermer et al.(1996)]{1996ApJ...456..106D} 
Dermer, C.~D., Miller, J.~A., \& Li, H.\ 1996, \apj, 456, 106
%
% Jet-Launching Structure Resolved Near the Supermassive Black Hole in M87
%\bibitem[Doeleman et al.(2012)]{ssd+12}
%Doeleman,~S.~S., et al., 2012, Science, 338, 355
%
%%%\bibitem[\protect\citeauthoryear{Dom{\'\i}nguez et al.}{2013}]{2013ApJ...770...77D} 
%%%Dom{\'\i}nguez A., Finke J.~D., Prada F., Primack J.~R., Kitaura F.~S., Siana B., Paneque D., 2013, ApJ, 770, 77

%
% Large-Amplitude X-Ray Outbursts from Galactic Nuclei: A Systematic Survey using ROSAT Archival Data
\bibitem[Donley et al.(2002)]{dbeb02}
Donley~J.~L., Brandt~W.~N., Eracleous~M., Boller~T., 2002, AJ, 124, 1308
%
% The tidal disruption of a star by a massive black hole
\bibitem[Evans \& Kochanek(1989)]{1989ApJ...346L..13E} 
Evans, C.~R., \& Kochanek, C.~S.\ 1989, \apjl, 346, L13 
%
%Higher-order spin effects in the dynamics of compact binaries I. Equations of motion
%\bibitem[Faye et al.(2007)]{gla07} 
%Faye,~G., Blanchet,~L., Buonanno,~A., 2007, arXiv:0605139
%
%\bibitem[Fragile \& Anninos(2005)]{fa05} 
%Fragile, P. C., Anninos, P., 2005, ApJ, 623, 347
%
%\bibitem[Fragile et al.(2007)]{popj07} 
%Fragile,~P.~C., Blaes,~O.~M., Anninos,~P., J.~D.~Salmonson., 2007, ApJ, 668, 417
%
%\bibitem[Frank \& Rees(1976)]{fr76}
%Frank, J., Rees, M. J. 1976, MNRAS, 176, 633
%
% Giant AGN Flares and Cosmic Ray Bursts
\bibitem[Farrar \& Gruzinov(2009)]{2009ApJ...693..329F} 
Farrar, G.~R., \& Gruzinov, A.\ 2009, \apj, 693, 329 
%
\bibitem[Farrar \& Piran(2014)]{2014arXiv1411.0704F} 
Farrar, G.~R., \& Piran, T.\ 2014, arXiv:1411.0704 
%
%Tidal Disruptions of Main-sequence Stars of Varying Mass and Age: Inferences from the Composition of the Fallback Material
%
\bibitem[Gallegos-Garcia et al.(2018)]{2018ApJ...857..109G} 
Gallegos-Garcia, M., Law-Smith, J., \& Ramirez-Ruiz, E.\ 2018, \apj, 857, 109
%
% 
% Ultraviolet Detection of the Tidal Disruption of a Star by a Supermassive Black Hole
\bibitem[Gezari et al.(2006)]{2006ApJ...653L..25G} 
Gezari, S., Martin, D.~C., Milliard, B., et al.\ 2006, \apjl, 653, L25
%
% UV/Optical Detections of Candidate Tidal Disruption Events by GALEX and CFHTLS
%%%%\bibitem[Gezari et al.(2008)]{2008ApJ...676..944G}
%%%Gezari, S., Basa, S., Martin, D.~C., et al.\ 2008, \apj, 676, 944-969 
%
\bibitem[Gezari et al.(2012)]{sg+12}
Gezari, S., Chornock, R., Rest, A., et al. 2012, Nature, 485, 217

%Gezari, S., Chornock, R., Lawrence, A., et al. 2015, ApJ, 815, L5
%Gezari, S., Martin, D. C., Milliard, B., et al. 2006, ApJ, 653, L25
%Gezari, S., Basa, S., Martin, D. C., et al. 2008, ApJ, 676, 944
%Gezari, S., Heckman, T., Cenko, S. B., et al. 2009, ApJ, 698, 1367
%Gezari, S., Hung, T., Cenko, S. B., et al. 2017, ApJ, 835, 144
%
% Near-infrared flares from accreting gas around the supermassive black hole at the galactic centre
%\bibitem[Genzel et al.(2003)]{rg+03}
%Genzel,~R., et al., 2003, Nature 425, 934
% 
%\bibitem[Giersz \& Spurzem(1994)]{gs94}
%Giersz, M., Spurzem, R. 1994, MNRAS, 269, 241
%
% Ultra-high-energy cosmic rays and neutrinos from tidal disruptions by massive black holes
\bibitem[Gu{\'e}pin et al.(2018)]{2018A&A...616A.179G} 
Gu{\'e}pin, C., Kotera, K., Barausse, E., et al.\ 2018, \aap, 616, A179.
%
%THREE-DIMENSIONAL SIMULATIONS OF TIDALLY DISRUPTED SOLAR-TYPE STARS AND THE OBSERVATIONAL SIGNATURES OF SHOCK BREAKOUT
%
\bibitem[Guillochon et al.(2009)]{2009ApJ...705..844G} 
Guillochon, J., Ramirez-Ruiz, E., Rosswog, S., \& Kasen, D.\ 2009, \apj, 705, 844 
%
%\bibitem[Guillochon \& Ramirez-Ruiz(2013)]{2013ApJ...767...25G} 
%Guillochon, J., \& Ramirez-Ruiz, E.\ 2013, \apj, 767, 25 
%
%\bibitem[Guillochon et al.(2014)]{gmr14} 
%Guillochon, J., Manukian, H., Ramirez-Ruiz, E., 2014, ApJ, 783, 23
%
\bibitem[Guillochon et al.(2016)]{2016ApJ...822...48G} 
Guillochon, J., McCourt, M., Chen, X., Johnson, M.~D., \& Berger, E.\ 2016, \apj, 822, 48
%
% Performance analysis of direct $N$-body algorithms on special-purpose supercomputers
%\bibitem[Harfst et al.(2007)]{2007NewA...12..357H} 
%Harfst, S., Gualandris, A., Merritt, D., et al.\ 2007, \na, 12, 357 
%
%\bibitem[Hayasaki et al.(2007)]{kh07}
%Hayasaki,~K., Mineshige,~S., Sudou,~H., 2007, PASJ, 59, 427
%
% Finite, intense accretion bursts from tidal disruption of stars on bound orbits
\bibitem[Hayasaki et al.(2013)]{2013MNRAS.434..909H}
Hayasaki,~K., Stone,~N., Loeb,~A. 2013, MNRAS, 434, 909
%
% Circularization of Tidally Disrupted Stars around Spinning Supermassive Black Holes
\bibitem[Hayasaki et al.(2016)]{2016MNRAS.461.3760H} 
Hayasaki, K., Stone, N., Loeb,~A. 2016, MNRAS, 461, 3760
%
\bibitem[Hayasaki et al.(2018)]{2018ApJ...855..129H} 
Hayasaki, K., Zhong, S., Li, S., et al.\ 2018, \apj, 855, 129.
%
%\bibitem[Heggie \& Mathieu(1986)]{hm86}
%Heggie, D. C., Mathieu, R. D., 1986, in Lecture Notes in Physics,
%Berlin Springer Verlag, Vol. 267, The Use of Supercomputers in
%Stellar Dynamics, ed. P. Hut \& S. L. W. McMillan, 233
%
%\bibitem[H{\'e}non(1971)]{1971Ap&SS..13..284H}
%H{\'e}non, M.\ 1971, \apss, 13, 284
%
% ASASSN-14ae: a tidal disruption event at 200 Mpc
\bibitem[Holoien et al.(2014)]{ho14}
Holoien, T. W.-S., Prieto, J. L., Bersier, D., et al. 2014, MNRAS, 445, 3263
%
\bibitem[Holoien et al.(2016)]{2016MNRAS.455.2918H} 
Holoien, T.~W.-S., Kochanek, C.~S., Prieto, J.~L., et al.\ 2016, \mnras, 455, 2918
%
\bibitem[Hung et al.(2017)]{2017ApJ...842...29H} 
Hung, T., Gezari, S., Blagorodnova, N., et al.\ 2017, \apj, 842, 29 
%
\bibitem[\protect\citeauthoryear{IceCube Collaboration, et al.}{2019}]{2019arXiv190205792I} 
IceCube Collaboration, et al., 2019, arXiv e-prints, arXiv:1902.05792
%
%\bibitem[Jiang et al.(2014)]{yanfei14}
%Jiang,~Y.-F., Stone,~J.~M., Davis,~S.~W., 2014, ApJ, submitted.
%
\bibitem[Jiang et al.(2016)]{2016ApJ...830..125J} 
Jiang, Y.-F., Guillochon, J., \& Loeb, A.\ 2016, \apj, 830, 125
%
\bibitem[\protect\citeauthoryear{Jiang, Stone \& Davis}{2017}]{2017arXiv170902845J} 
Jiang Y.-F., Stone J., Davis S.~W., 2017, arXiv e-prints, arXiv:1709.02845
%
%\bibitem[Kato et al.(2008)]{sjs08} S.~Kato, J.~Fukue, S.~Mineshige, 
%(Kyoto: Kyoto University Press), Black-Hole Accretion Disks Towards a New Paradigm (2008)
%
% Enhanced Accretion Rates of Stars on Supermassive Black Holes by Star-Disk Interactions in Galactic Nuclei
%\bibitem[Just et al.(2012)]{2012ApJ...758...51J} 
%Just, A., Yurin, D., Makukov, M., et al.\ 2012, \apj, 758, 51
%
% Optical Transients from the Unbound Debris of Tidal Disruption
%\bibitem[Kasen \& Ramirez-Ruiz(2010)]{2010ApJ...714..155K} 
%Kasen, D., \& Ramirez-Ruiz, E.\ 2010, \apj, 714, 155 
%
%
\bibitem[Kelner et al.(2006)]{2006PhRvD..74c4018K} 
Kelner, S.~R., Aharonian, F.~A., \& Bugayov, V.~V.\ 2006, \prd, 74, 034018 
%
%
% Star-disc interaction in galactic nuclei: orbits and rates of accreted stars
%\bibitem[Kennedy et al.(2016)]{2016MNRAS.460..240K} 
%Kennedy, G.~F., Meiron, Y., Shukirgaliyev, B., et al.\ 2016, \mnras, 460, 240 
%
%\bibitem[Khan et al.(2012)]{khan+12}
%Khan, F. M., Berentzen, I., Berczik, P., Just, A., Mayer, L., Nitadori, K., Callegari, S. 2012, ApJ, 756, 30
%
%\bibitem[Khan et al.(2016)]{khan+16}
%Khan, F. M., Fiacconi, D., Mayer, L., Berczik, P., Just, A. 2016, ApJ, 828, 73
%
% Neutrino and Cosmic-Ray Emission and Cumulative Background from Radiatively Inefficient Accretion Flows in Low-luminosity Active Galactic Nuclei
\bibitem[Kimura et al.(2015)]{2015ApJ...806..159K} 
Kimura, S.~S., Murase, K., \& Toma, K.\ 2015, \apj, 806, 159 
%
% Gravitational Waves and X-Ray Signals from Stellar Disruption by a Massive Black Hole
\bibitem[Kobayashi et al.(2004)]{2004ApJ...615..855K} 
Kobayashi, S., Laguna, P., Phinney, E.~S., \& M{\'e}sz{\'a}ros, P.\ 2004, \apj, 615, 855
%
%
%\bibitem[Kochanek(1994)]{kochanek94} 
%Kochanek, C.~S., 1994, ApJ, 422, 508
%
%
% Abundance anomalies in tidal disruption events
%
\bibitem[Kochanek(2016)]{2016MNRAS.458..127K} 
Kochanek, C.~S.\ 2016, \mnras, 458, 127 
%
%
%Tidal disruption event demographics
%
\bibitem[Kochanek(2016)]{2016MNRAS.461..371K} 
Kochanek, C.~S.\ 2016, \mnras, 461, 371
%
% Extraction of Black Hole Rotational Energy by a Magnetic Field and the Formation of Relativistic Jets
%\bibitem[Koide et al.(2002)]{sktd02} 
%Koide,~S., Shibata,~K., Kudoh,~T., Meier,~D.~L., 2002, Science 295, 1688
%
%
% The giant X-ray outbursts in NGC 5905 and IC 3599: Follow-up observations and outburst scenarios
%\bibitem{kg99}  S.~Komossa, J.~Greiner, Astron. Astrophys. 349, L45 (1999).
% The giant X-ray outbursts in NGC 5905 and IC 3599: Follow-up observations and outburst scenarios
\bibitem[Komossa \& Bade(1999)]{kb99}
Komossa~S., Bade~N., 1999, A\&A, 343, 775
%
%%%\bibitem[Komossa(2015)]{2015JHEAp...7..148K} 
%%%Komossa, S.\ 2015, Journal of High Energy Astrophysics, 7, 148 
%
% Inward bound\UTF{0081}\The search for supermassive black holes in galactic nuclei.
%\bibitem[Kormendy \& Richstone(1995)]{kr95}
%Kormendy,~J., Richstone,~D., 1995, ARA\&A, 33, 581
%
%Coevolution (Or Not) of Supermassive Black Holes and Host Galaxies
%\bibitem[Kormendy \& Ho(2013)]{kho13}
%Kormendy,~J., Ho,~L.~C., 2013, ARA\&A, 51, 511
%
% Stellar dynamics in the galactic centre: Proper motions and anisotropy
%\bibitem{rg+00} R.~Genzel., et al., Mon. Not. R. Astron. Soc. 317, 348 (2000).
%
%
%\bibitem[Laguna et al.(1993a)]{laguna93a}
%Laguna P., Miller W. A., Zurek W. H., 1993a, ApJ, 404, 678
%
%\bibitem[Laguna et al.(1993b)]{laguna93b}
%Laguna P., Miller W. A., Zurek W. H., Davies M. B., 1993b, ApJ, 410, L83
%
%\bibitem[Li et al.(2017)]{2017ApJ...834..195L} 
%Li, S., Liu, F.~K., Berczik, P., \& Spurzem, R.\ 2017, \apj, 834, 195 
%
% An Extremely Luminous Panchromatic Outburst from the Nucleus of a Distant Galaxy
%\bibitem[Levan et al.(2011)]{ajl+11} 
%Levan,~A.~J. et al., 2011, Science, 333, 199
%
% Swift discoveries of new populations of extremely long duration high energy transient
%%%\bibitem[Levan(2015)]{2015JHEAp...7...44L} 
%%%Levan, A.~J.\ 2015, Journal of High Energy Astrophysics, 7, 44 
%
%
%\bibitem[Lodato et al.(2009)]{lkp09}
%Lodato, G., King, A.R., Pringle, J.E. 2009, 392, 332
\bibitem[Lodato et al.(2009)]{2009MNRAS.392..332L} 
Lodato, G., King, A.~R., \& Pringle, J.~E.\ 2009, \mnras, 392, 332
%
% Multiband light curves of tidal disruption events
%\bibitem[Lodato \& Rossi(2011)]{lr11}
%Lodato G., Rossi E. M., 2011, MNRAS, 410, 359
%
%OPTICAL APPEARANCE OF THE DEBRIS OF A STAR DISRUPTED BY A MASSIVE
%\bibitem[Loeb \& Ulmer(1997)]{lu97}
%Loeb,~A., Ulmer,~A., 1997, ApJ, 489, 573
%
%High energy neutrinos from the tidal disruption of stars
\bibitem[Lunardini \& Winter(2017)]{2017PhRvD..95l3001L} 
Lunardini, C., \& Winter, W.\ 2017, \prd, 95, 123001 
%
% The Demography of Massive Dark Objects in Galaxy Centers
%\bibitem[Magorrian et al.(1998)]{mtr98} 
%Magorrian, J., Tremaine, S., Richstone,~D., et al. 1998, AJ 115, 2285
%
%\bibitem{hills75} J.~G.~Hills, Nature 254, 295 (1975).
%
% Energy input from quasars as a way to regulate the growth of of black holes and host galaxies
%\bibitem{ms05} T.~Di Matteo, V.~Springel, L. Hernquist, Nature 433, 604 (2005).
%
%
%\bibitem{mp04} D.~Merritt, M.~Y.~Poon, Astrophys. J. 606, 788 (2004).
% Rates of tidal disruption of stars by massive central black holes
%\bibitem[Magorrian \& Tremaine(1999)]{mt99} 
%Magorrian, J., Tremaine, S., 1999, MNRAS, 309, 447 
%
% Scaling Laws for Advection-dominated Flows: Applications to Low-Luminosity Galactic Nuclei
\bibitem[Mahadevan(1997)]{1997ApJ...477..585M} 
Mahadevan, R.\ 1997, \apj, 477, 585 
%
%
\bibitem[Maksym et al.(2013)]{2013MNRAS.435.1904M} 
Maksym, W.~P., Ulmer, M.~P., Eracleous, M.~C., Guennou, L., \& Ho, L.~C.\ 2013, \mnras, 435, 1904
%
%
\bibitem[\protect\citeauthoryear{Marshall, Avara \& McKinney}{2018}]{2018MNRAS.478.1837M} 
Marshall M.~D., Avara M.~J., McKinney J.~C., 2018, MNRAS, 478, 1837
%
%
\bibitem[\protect\citeauthoryear{McKinney, Tchekhovskoy \& Blandford}{2012}]{2012MNRAS.423.3083M} 
McKinney J.~C., Tchekhovskoy A., Blandford R.~D., 2012, MNRAS, 423, 3083
%
\bibitem[\protect\citeauthoryear{McKinney, Dai \& Avara}{2015}]{2015MNRAS.454L...6M} 
McKinney J.~C., Dai L., Avara M.~J., 2015, MNRAS, 454, L6
%
%
%Revisiting the scaling relations of black hole masses and host galaxy properties?
%\bibitem[McConnell \& Ma (2013)]{mm13} 
%McConnell, N. J., Ma, C.-P., 2013, \apj, 764, 184
%
% Black Hole Spin via Continuum Fitting and the Role of Spin in Powering Transient Jets
%\bibitem[McClintock et al.(2014)]{Mc14}
%McClintock, J. E., Narayan, R., \& Steiner, J. F. 2014, SSRv, 183, 295
%
% Alignment of Magnetized Accretion Disks and Relativistic Jets with Spinning Black Holes
%\bibitem[McKinney et al.(2013)]{jar13}
%McKinney,~J.~C., Tchekhovskoy,~A., Blandford,~R.~D., Science, 2013, 339, 49
%
% Astrophysical Sources of High-Energy Neutrinos in the IceCube Era
\bibitem[M{\'e}sz{\'a}ros(2017)]{2017ARNPS..67...45M} 
M{\'e}sz{\'a}ros, P.\ 2017, Annual Review of Nuclear and Particle Science, 67, 45 
%
%Testing properties of the Galactic center black hole using stellar orbits
\bibitem[Merritt et al.(2010)]{2010PhRvD..81f2002M} 
Merritt, D., Alexander, T., Mikkola, S., \& Will, C.~M.\ 2010, \prd, 81, 062002 
%\bibitem[Merritt et al.(2011)]{mamw11} 
%Merritt, D., Alexander, T., Mikkola, S., \& Will, C.~M.\ 2011, PRD, 84, 044024 
%%%\bibitem[Merritt(2013)]{2013CQGra..30x4005M} 
%%%Merritt, D.\ 2013, Classical and Quantum Gravity, 30, 244005
% 
% The SDSS DR6 luminosity functions of galaxies
%%%\bibitem[Montero-Dorta \& Prada(2009)]{2009MNRAS.399.1106M} 
%%%Montero-Dorta, A.~D., \& Prada, F.\ 2009, \mnras, 399, 1106
%
% Gravitational-wave sensitivity curves
\bibitem[\protect\citeauthoryear{Moore, Cole \& Berry}{2015}]{2015CQGra..32a5014M} 
Moore C.~J., Cole R.~H., Berry C.~P.~L., 2015, CQGra, 32, 15014
%
% Astrophysical high-energy neutrinos and gamma-ray bursts
%\bibitem[Murase(2008)]{2008AIPC.1065..201M} 
%Murase, K.\ 2008, American Institute of Physics Conference Series, 1065, 201 
\bibitem[Murase(2008)]{2008AIPC.1065..201M} 
Murase, K.\ 2008, American Institute of Physics Conference Series, 201
%
% Energetics of high-energy cosmic radiations
\bibitem[Murase \& Fukugita(2019)]{2019PhRvD..99f3012M} 
Murase, K., \& Fukugita, M.\ 2019, \prd, 99, 63012.
%
% Hidden Cosmic-Ray Accelerators as an Origin of TeV-PeV Cosmic Neutrinos
\bibitem[Murase et al.(2016)]{2016PhRvL.116g1101M} 
Murase, K., Guetta, D., \& Ahlers, M.\ 2016, Physical Review Letters, 116, 071101 
%
% TeV-PeV Neutrinos from Low-Power Gamma-Ray Burst Jets inside Stars
\bibitem[Murase \& Ioka(2013)]{2013PhRvL.111l1102M}
Murase, K., \& Ioka, K.\ 2013, Physical Review Letters, 111, 121102 
%
\bibitem[\protect\citeauthoryear{Narayan, Igumenshchev \& Abramowicz}{2003}]{2003PASJ...55L..69N} 
Narayan R., Igumenshchev I.~V., Abramowicz M.~A., 2003, PASJ, 55, L69
%
% Advection-dominated accretion: A self-similar solution
\bibitem[Narayan \& Yi(1994)]{1994ApJ...428L..13N} 
Narayan, R., \& Yi, I.\ 1994, \apjl, 428, L13 
%
% Advection-dominated Accretion: Underfed Black Holes and Neutron Stars
\bibitem[Narayan \& Yi(1995)]{1995ApJ...452..710N} 
Narayan, R., \& Yi, I.\ 1995, \apj, 452, 710 
%
%\bibitem[Mihalas \& Mihalas(1984)]{mm84}
%Mihalas, D., Mihalas, B. W. 1984, Foundations of radiation hydrodynamics, ed. Mihalas, D. \& Mihalas, B. W.
%
% Optical flares from the tidal disruption of stars by massive black holes
%\bibitem{sq09} L.~E.~Strubbe, E.~Quataert, Mon. Not. R. Astron. Soc. 400, 2070 (2009).
%
% Supercritical Accretion Flows around Black Holes: Two-dimensional, Radiation Pressure-dominated Disks with Photon Trapping
%\bibitem{omnm05} K.~Ohsuga, M.~Mori, T.~Nakamoto, S.~Mineshige, Astrophys.~J. 628, 368 (2005).
%
% XMM-EPIC observation of MCG?6-30-15: direct evidence for the extraction of energy from a spinning black hole?
%\bibitem{jw01} J. Wilms et al., Mon. Not. R. Astron. Soc. 328, L27 (2001).
%
% Relativistic X-Ray Lines from the Inner Accretion Disks Around Black Holes
%\bibitem[Miller(2007)]{miller07} 
%Miller,~J.~M., 2007, ARA\&A, 45, 441
%
% Evidence for a black hole from high rotation velocities in a sub-parsec region of NGC4258,
%\bibitem[Miyoshi et al.(1995)]{mm+95} 
%Miyoshi,~M., et al., 1995, Nature 373, 127
%
%\bibitem[Nelson \& Papaloizou(2000)]{np00}
%Nelson, R. P., Papaloizou, J. C. B., 2000, MNRAS, 315, 570
%
% Two-dimensional, Radiation Pressure-dominated Disks with Photon Trapping
%\bibitem[Ohsuga et al.(2005)]{ohsuga05}
%Ohsuga,~K., Mori,~M., Nakamoto,~T., Mineshige,~S., 2005, ApJ, 628, 368
%
%\bibitem[Okazaki et al.(2002)]{ato02}
%Okazaki A.T., Bate M.R., Ogilvie G.I., Pringle J.E., 2002, MNRAS, 337, 967
%
\bibitem[Phinney(1989)]{p89}
Phinney, E. S. 1989, in IAU Symp. 136, The Center of the Galaxy, ed.
M. Morris (Dordrecht: Kluwer Academic Publishers), 543
%
%
\bibitem[Pfeffer et al.(2017)]{2017MNRAS.466.2922P} 
Pfeffer, D.~N., Kovetz, E.~D., \& Kamionkowski, M.\ 2017, \mnras, 466, 2922
%
% The Star Ingesting Luminosity of Intermediate-Mass Black Holes in Globular Clusters
%\bibitem[Ramirez \& Rosswog(2009)]{rr09}
%Ramirez-Ruiz,~E., Rosswog,~S., 2009, ApJ, 697, 77
%
% Disk Formation Versus Disk Accretion?What Powers Tidal Disruption Events?
\bibitem[Piran et al.(2015)]{2015ApJ...806..164P} 
Piran, T., Svirski, G., Krolik, J., Cheng, R.~M., \& Shiokawa, H.\ 2015, \apj, 806, 164 
%
%
\bibitem[Rees(1988)]{rees88} 
Rees,~M.~J., 1998, Nature 333, 523 
%
%
% A rapidly spinning supermassive black hole at the centre of NGC 1365
%\bibitem[Risaliti et al.(2013)]{gr+13} 
%Risaliti,~G., et al., 2013, Nature 452, 449
%
\bibitem[Saxton et al.(2012)]{2012A&A...541A.106S} 
Saxton, R.~D., Read, A.~M., Esquej, P., et al.\ 2012, \aap, 541, A106
%
% A star in a 15.2 year orbit around the supermassive black hole at the centre of the Milky Way
%\bibitem[Sch$\ddot{\rm o}$del et al(2002)]{rs+02} 
%Sch$\ddot{\rm o}$del,~R., et al, 2002, Nature, 419, 694
%
%\bibitem[Schulze\&Gebhardt(2011)]{sg11}
%Schulze, A., Gebhardt, K. 2011, ApJ, 729, 21
%
%
% High-energy Neutrino Flares from X-Ray Bright and Dark Tidal Disruption Events
\bibitem[Senno et al.(2017)]{2017ApJ...838....3S} 
Senno, N., Murase, K., \& M{\'e}sz{\'a}ros, P.\ 2017, \apj, 838, 3.
%
%
\bibitem[Shiokawa et al.(2015)]{2015ApJ...804...85S} 
Shiokawa, H., Krolik, J.~H., Cheng, R.~M., Piran, T., \& Noble, S.~C.\ 2015, \apj, 804, 85 
%
%\bibitem[Shakura \& Sunyaev(1973)]{ss73}
%Shakura,~N.~I., Sunyaev,~R.~A., 1973, A\&A, 24, 337
%
%
%\bibitem[Sigurdsson \& Rees(1997)]{1997MNRAS.284..318S} 
%Sigurdsson, S., \& Rees, M.~J.\ 1997, \mnras, 284, 318
%
%
%\bibitem[Spurzem et al.(2012)]{spur12}
%Spurzem, R., Berczik, P., Zhong, S., Nitadori, K., Hamada, T., Berentzen, I., Veles, A. 2012, in Astronomical Society of the Pacific Conference Series, Vol. 453, Advances in Computational Astrophysics: Methods, Tools, and Outcome, ed. R. Capuzzo-Dolcetta, M. Limongi, and A. Tornamb\'e, 223
%
% Observing Lense-Thirring Precession in Tidal Disruption Flares
%\bibitem[Stone \& Loeb(2012a)]{na12} 
%Stone, N., Loeb, A., 2012, Physical Review Letters, 108, 061302
%
\bibitem[Stone et al.(2013)]{nra13} 
Stone, N., Sari, R., Loeb, A., 2013, MNRAS, 435, 1809 
%
% Rates of Stellar Tidal Disruption as Probes of the Supermassive Black Hole Mass Function
\bibitem[Stone \& Metzger(2016)]{sm16}
Stone, N., Metzger, B. D. 2016, MNRAS, 455, 859
%
%\bibitem[Stone et al.(2015)]{nka15} 
%Stone,~N., Hayasaki,~K., Loeb,~A., 2015, in prep.
%Physical Review Letters, submitted.
%
% Optical flares from the tidal disruption of stars by massive black holes
%%%\bibitem[Strubbe \& Quataert(2009)]{2009MNRAS.400.2070S} 
%%%Strubbe, L.~E., \& Quataert, E.\ 2009, \mnras, 400, 2070 
%
%Spectroscopic signatures of the tidal disruption of stars by massive black holes
%\bibitem[Strubbe \& Quataert(2011)]{sq11} 
%Strubbe,~L.~E., Quataert,~E., 2011, MNRAS, 415, 168
%
%\bibitem[Strubbe \& Murray(2015)]{2015MNRAS.454.2321S} 
%Strubbe, L.~E., \& Murray, N.\ 2015, \mnras, 454, 2321

%
% On the Momentum Diffusion of Radiating Ultrarelativistic Electrons in a Turbulent Magnetic Field
%
\bibitem[Stawarz \& Petrosian(2008)]{2008ApJ...681.1725S} 
Stawarz, {\L}., \& Petrosian, V.\ 2008, \apj, 681, 1725
%
%
% Gravitationally redshifted emission implying an accretion disk and massive black hole in the active galaxy MCG-6-30-15
%\bibitem[Tanaka et al.(1995)]{yt+95} 
%Tanaka,~Y., et al., 1995, Nature 375, 659
%
%
% Prograde and retrograde black holes: Whose jet is more powerful? 
%\bibitem[Tchekhovskoy et al.(2012)]{ajc12}
%Tchekhovskoy,~A., McKinney,~J.~C., 2012, MNRAS, 423, L55
%
% 
% Magnetohydrodynamics in the extreme relativistic limit
\bibitem[Thompson \& Blaes(1998)]{1998PhRvD..57.3219T} 
Thompson, C., \& Blaes, O.\ 1998, \prd, 57, 3219 
%
\bibitem[van Velzen \& Farrar(2014)]{vf14} 
van Velzen, S., Farrar, G.~R., 2014, ApJ, 792, 53 
%
\bibitem[van Velzen et al.(2016)]{2016Sci...351...62V} 
van Velzen, S., Anderson, G.~E., Stone, N.~C., et al.\ 2016, Science, 351, 62 
%
%A Luminous, Fast Rising UV-transient Discovered by ROTSE: A Tidal Disruption Event?
%\bibitem[Vink\'o et al.(2015)]{vi+15} Vink\'o,~J., et al., 2015, ApJ, 798, 12
\bibitem[Vink{\'o} et al.(2015)]{2015ApJ...798...12V}
Vink{\'o}, J., Yuan, F., Quimby, R.~M., et al.\ 2015, \apj, 798, 12 
%
%\bibitem[Volonteri et al.(2005)]{vmq+05} 
%Volonteri, M., Madau, P., Quataert, E., Rees, M.~J., 2005, ApJ, 620, 69 
%
%
%Revised Rates of Stellar Disruption in Galactic Nuclei
\bibitem[Wang \& Merritt(2004)]{jd04} 
Wang,~J., Merritt,~D., 2004, ApJ, 600, 149
%
% Pseudo-Newtonian Potentials for Nearly Parabolic Orbits
%\bibitem[Wegg(2012)]{wc12} 
%Wegg,~C., 2012, ApJ, 749, 183
%
% Probing the tidal disruption flares of massive black holes with high-energy neutrinos
\bibitem[Wang et al.(2011)]{2011PhRvD..84h1301W} 
Wang, X.-Y., Liu, R.-Y., Dai, Z.-G., \& Cheng, K.~S.\ 2011, \prd, 84, 081301 
%
% Tidal disruption jets of supermassive black holes as hidden sources of cosmic rays: Explaining the IceCube TeV-PeV neutrinos
\bibitem[Wang \& Liu(2016)]{2016PhRvD..93h3005W} 
Wang, X.-Y., \& Liu, R.-Y.\ 2016, \prd, 93, 083005 
% 
% Self-similar Solution of Optically Thick Advection-dominated Flows
\bibitem[Wang \& Zhou(1999)]{1999ApJ...516..420W} 
Wang, J.-M., \& Zhou, Y.-Y.\ 1999, \apj, 516, 420 
%
\bibitem[\protect\citeauthoryear{Watarai}{2006}]{2006ApJ...648..523W} 
Watarai K.-. ya ., 2006, ApJ, 648, 523
%
%
\bibitem[Waxman(1995)]{1995PhRvL..75..386W} 
Waxman, E.\ 1995, \prl, 75, 386.
%
%
\bibitem[Waxman \& Loeb(2001)]{2001PhRvL..87g1101W} 
Waxman, E., \& Loeb, A.\ 2001, Physical Review Letters, 87, 071101 
%
%
%\bibitem[\protect\citeauthoryear{Yamazaki, Kohri \& Katagiri}{2009}]{2009A&A...495....9Y} 
%Yamazaki R., Kohri K., Katagiri H., 2009, A\&A, 495, 9
%
% Birth of a relativistic outflow in the unusual \UTF{0083}\UTF{00C1}-ray transient Swift J164449.3+573451
\bibitem[Zauderer et al.(2011)]{baz+11}
Zauderer, B. A., Berger, E., Soderberg, A. M., et al. 2011, Nature, 476, 425
%High-energy cosmic ray nuclei from tidal disruption events: Origin, survival, and implications
\bibitem[Zhang et al.(2017)]{2017PhRvD..96f3007Z} 
Zhang, B.~T., Murase, K., Oikonomou, F., \& Li, Z.\ 2017, \prd, 96, 063007 
%
%\bibitem[Zhong et al.(2014)]{zhong+14}
%Zhong, S., Berczik, P., Spurzem, R. 2014, ApJ, 792, 137
%
%\bibitem[Zhong et al.(2018)]{zhong+18}
%Zhong, S., Hayasaki, K., Li, S., et al. 2018, in prep.
%Berczik, P., Spurzem, R. 2014, ApJ, 792, 137
%
\end{thebibliography}
\end{document}